\pdfoutput=1

\documentclass[twoside,journal]{IEEEtran}

\usepackage{color}
\usepackage{cite}
\usepackage{amsmath}
\usepackage{amssymb}
\usepackage{graphicx}
\usepackage{epstopdf}
\usepackage{tabularx}
\usepackage{booktabs} 
\usepackage[usenames,dvipsnames]{xcolor}
\usepackage{eso-pic}
\usepackage{textcomp}
\usepackage{pifont}
\usepackage{anyfontsize}
\usepackage{changepage}
\usepackage{setspace}
\usepackage{url}
\usepackage{multirow}

\ifCLASSINFOpdf
\else
\fi

\IEEEoverridecommandlockouts

\AddToShipoutPicture*{\small \raisebox{0.9cm}
{\hspace*{2.2cm}
\begin{minipage}[c]{0.95\textwidth}
\footnotesize
{\setlength{\baselineskip}{0.5\baselineskip}
\textcopyright 2018 IEEE. Personal use of this material is permitted. Permission from IEEE must be obtained for all other uses, including reprinting/republishing this material for advertising or promotional purposes, collecting new collected works for resale or redistribution to servers or lists, or reuse of any copyrighted component of this work in other works.
\par}
\end{minipage}}}%

\AddToShipoutPicture*{\small \raisebox{0.35cm}
{\hspace*{-17.33cm}
\begin{minipage}[c]{0.95\textwidth}
\footnotesize
{\setlength{\baselineskip}{0.5\baselineskip}
This document is the paper as accepted for publication in TBioCAS, the fully-edited paper is available at \url{https://ieeexplore.ieee.org/document/8528875}.
\par}
\end{minipage}}}%

\AddToShipoutPicture*{\small \raisebox{0.08cm}
{\hspace*{-17.33cm}
\begin{minipage}[c]{0.95\textwidth}
\footnotesize
{\setlength{\baselineskip}{0.5\baselineskip}
DOI: 10.1109/TBCAS.2018.2880425
\par}
\end{minipage}}}%

\begin{document}

\title{\vspace*{-1.5mm}A 0.086-mm$^2$ 12.7-pJ/SOP 64k-Synapse 256-Neuron Online-Learning Digital Spiking Neuromorphic Processor in 28nm CMOS}
\author{Charlotte~Frenkel,~\IEEEmembership{Student Member,~IEEE,}
		Martin~Lefebvre,~\IEEEmembership{Student Member,~IEEE,}\\
        Jean-Didier~Legat,~\IEEEmembership{Senior Member,~IEEE,}
        and~David~Bol,~\IEEEmembership{Senior Member,~IEEE}\vspace*{-2mm}%
\thanks{Manuscript received xxxxx xx, 2018; revised xxxxx xx, 2018; accepted xxxxx xx, 2018. Date of publication xxxxx xx, 2018; date of current version xxxxx xx, 2018.}%
\thanks{The authors are with the ICTEAM Institute, Universit\'e catholique de Louvain, Louvain-la-Neuve BE-1348, Belgium (e-mail: \{charlotte.frenkel, martin.lefebvre, jean-didier.legat, david.bol\}@uclouvain.be).}%
\thanks{C. Frenkel is with Universit\'e catholique de Louvain as a Research Fellow from the National Foundation for Scientific Research (FNRS) of Belgium.}%
\thanks{Color versions of one or more of the figures in this paper are available online at http://ieeexplore.ieee.org.}%
\thanks{Digital Object Identifier xx.xxxx/TBCAS.2018.xxxxxxx}}%

% The paper headers
\markboth{IEEE TRANSACTIONS ON BIOMEDICAL CIRCUITS AND SYSTEMS,~VOL.~xx, NO.~xx, XXXXX~2018}%
{FRENKEL \MakeLowercase{\textit{et al.}}: A 0.086-\MakeLowercase{mm}$^2$ 12.7-\MakeLowercase{p}J/SOP 64\MakeLowercase{k}-SYNAPSE 256-NEURON ONLINE-LEARNING DIGITAL SPIKING NEUROMORPHIC PROCESSOR}

\maketitle

\begin{abstract}
Shifting computing architectures from von Neumann to event-based spiking neural networks (SNNs) uncovers new opportunities for low-power processing of sensory data in applications such as vision or sensorimotor control. Exploring roads toward cognitive SNNs requires the design of compact, low-power and versatile experimentation platforms with the key requirement of online learning in order to adapt and learn new features in uncontrolled environments. However, embedding online learning in SNNs is currently hindered by high incurred complexity and area overheads. In this work, we present ODIN, a \mbox{0.086-mm$^2$} 64k-synapse 256-neuron online-learning digital spiking neuromorphic processor in 28nm FDSOI CMOS achieving a minimum energy per synaptic operation (SOP) of 12.7pJ. It leverages an efficient implementation of the spike-driven synaptic plasticity (SDSP) learning rule for high-density embedded online learning with only 0.68$\mu$m$^2$ per 4-bit synapse. Neurons can be independently configured as a standard leaky integrate-and-fire (LIF) model or as a custom phenomenological model that emulates the 20 Izhikevich behaviors found in biological spiking neurons. Using a single presentation of 6k 16$\times$16 MNIST training images to a single-layer fully-connected 10-neuron network with on-chip SDSP-based learning, ODIN achieves a classification accuracy of 84.5\% while consuming only 15nJ/inference at 0.55V using rank order coding. ODIN thus enables further developments toward cognitive neuromorphic devices for low-power, adaptive and low-cost processing.

\end{abstract}


\IEEEpeerreviewmaketitle

\section{Introduction} \label{sec_intro}

\IEEEPARstart{W}{hile} a massive deployment of the Internet-of-Things (IoT) paradigm within the upcoming years sets stringent requirements for autonomous smart sensors design~\cite{Bol15}, the end of Moore's law~\cite{Horowitz14} calls for new computing architectures that can accommodate severe cost and power reduction constraints. In radical contrast with current von Neumann processing architectures, biological brains appear to have an unmatched performance-resource tradeoff~\cite{Sandin14}: for example, the bee brain has close to 1 million neurons and a power consumption around 10$\mu$W, yet it is capable of complex behaviors such as sequence and pattern learning, navigation, planning and anticipation, while exhibiting a learning speed outperforming human infants~\cite{Chittka09,Liu10}. Therefore, in order to bring silicon information processing devices closer to the efficiency of biological brains, the field of neuromorphic engineering addresses the study and design of bio-inspired systems following a paradigm shift along two axes. The first axis is linked to computation organization: biological neural networks feature co-located processing (i.e.~neurons) and memory (i.e.~synapses) with massively-parallel data handling~\cite{Indiveri15a}. The second axis is linked to information representation: biological neural networks process information in the time domain, using spikes to encode data. Information processing is entirely event-driven, leading to sparse low-power computation~\cite{Liu10}.

This two-fold paradigm shift could lead to new bio-inspired and power-efficient neuromorphic computing devices, whose sparse event-driven data acquisition and processing appear to be particularly suited for distributed autonomous smart sensors for the IoT relying on energy harvesting~\cite{Bol15,Sandin14}, closed sensorimotor loops for autonomous embedded systems and robots with strict battery requirements~\cite{Sandamirskaya14,Milde17}, brain-machine interfaces~\cite{Corradi15,Boi16} and neuroscience experimentation or bio-hybrid platforms~\cite{Vogelstein08,George15}. However, despite recent advances in neuroscience, detailed understanding of the computing and operating principles of the brain is still out of reach~\cite{Indiveri09}. This highlights the need for efficient experimentation platforms with high versatility in neuronal behaviors~\cite{Izhikevich04} and online learning with synaptic plasticity~\cite{Azghadi14} to explore brain-like computation toward efficient event-based SNN processors. A software approach has been proposed in~\cite{Painkras13} with SpiNNaker for large-scale neural network simulation, but high flexibility is achieved at the expense of limited power and area efficiencies. Similar conclusions hold for FPGA-based approaches (e.g.,~\cite{Cassidy13,Neil14,Luo16,Yang18}). Therefore, the challenge of low-power low-area large-scale integration of biophysically-accurate neuromorphic SNN devices needs to be overcome~\cite{Poon11}. 

The first design approach for neuromorphic devices appeared in the late 1980s and exploited direct emulation of the brain ion channels dynamics with the MOS transistor operated in subthreshold regime~\cite{Mead89}, an approach still popular today for SNNs~(e.g., ROLLS~\cite{Qiao15} and DYNAPs~\cite{Moradi17}). While subthreshold analog approaches emulate the brain dynamics with biological time constant for real-time sensorimotor control, above-threshold analog approaches allow simulating neurons with acceleration factors of up to 100,000 for fast processing~(e.g., BrainScaleS~\cite{Schemmel10}). A switched-capacitor analog implementation has also been proposed to ease robust analog design in deep submicron technologies~\cite{Noack15,Mayr16}. However, in order to fully leverage technology scaling, several research groups recently started designing digital SNNs~(e.g.,~Seo \textit{et al.} in~\cite{Seo11}, Kim \textit{et al.} in~\cite{Kim15}, IBM with TrueNorth~\cite{Akopyan15} and Intel with Loihi~\cite{Davies18}). Digital designs have a shorter design cycle, low sensitivity to noise, process-voltage-temperature (PVT) variations and mismatch, and suppress the need to generate bias currents and voltages. Depending on their implementation, digital SNNs can span biological to accelerated time constants and exhibit one-to-one correspondence between the fabricated~\mbox{hardware and the software model.}

The implementation of resource-efficient biophysically-accurate and versatile digital SNNs is still an open challenge as emulation of the brain dynamics requires the implementation of high-complexity neuron and synapse models. Indeed, two key ingredients are required. First, event-driven embedded online learning allows low-power autonomous agents to adapt in real time to new features in uncontrolled environments, where limited training data is presented on-the-fly to the network. These requirements cannot be met by conventional offline learning techniques in backpropagation-based artificial neural networks (ANNs) as they rely on repeated presentations of extensive training data. As the biological fan-in is on the order of 100 to 10,000 synapses per neuron, embedding locally an online learning rule such as spike-timing-dependent plasticity (STDP)~\cite{Markram12} or spike-driven synaptic plasticity (SDSP)~\cite{Brader07} in each single synapse is challenging~\cite{Frenkel17a}. Memristors promise new records, but high-yield co-integration with CMOS is still to be demonstrated~\cite{Lin14,Rofeh15}. Second, the widespread leaky integrate-and-fire (LIF) neuron model has been shown to lack the essential behavior repertoire necessary to explore the computational properties of large neural networks~\cite{Indiveri10}. In contrast, implementing biophysically-accurate models (e.g., Hodgkin-Huxley~\cite{Hodgkin52}, Izhikevich~\cite{Izhikevich03} or adaptive-exponential~\cite{Brette05}) requires digital SNNs to solve coupled non-linear differential equations and to update all neuron states at each integration timestep. Therefore, in this work, we propose ODIN, an \underline{\smash{o}}nline-learning \underline{\smash{di}}gital spiking \underline{\smash{n}}euromorphic processor in 28nm FDSOI CMOS. It comprises 256 neurons and 64k synapses and embeds SDSP online learning at a high density with only 0.68$\mu$m$^2$ per synapse. Neurons can be programmed to emulate all the 20 Izhikevich behaviors found in biological spiking neurons~\cite{Izhikevich04} using a custom phenomenological model that is entirely event-driven and does not require a neuron update at each timestep. ODIN occupies only 0.086mm$^2$ and consumes a minimum energy per synaptic operation~(SOP) of 12.7pJ at 0.55V. Using a single presentation of 6k 16$\times$16-pixel training images from the MNIST dataset of handwritten digits~\cite{LeCun98}, a single-layer fully-connected 10-neuron network with on-chip SDSP-based learning achieves a classification accuracy of 84.5\% with only 15nJ per inference using rank order coding.

The remainder of the paper is structured as follows. First, the design of ODIN is described in Section~\ref{sec_arch}, with architecture and implementation details on the event-based addressing scheme, the online-learning synapse, the phenomenological neuron model and the internal event scheduler. Second, specifications and measurement results are presented in Section~\ref{sec_mesu} with a comparison on MNIST of two learning strategies: on-chip and online with SDSP or off-chip and offline with stochastic gradient descent. Finally, these results are discussed and compared with the state of the art in Section~\ref{sec_disc}, before summarizing concluding remarks in Section~\ref{sec_ccl}.

\section{Architecture and Implementation} \label{sec_arch}

\begin{figure}[!t]
\centering
\noindent\includegraphics[width=0.90\columnwidth]{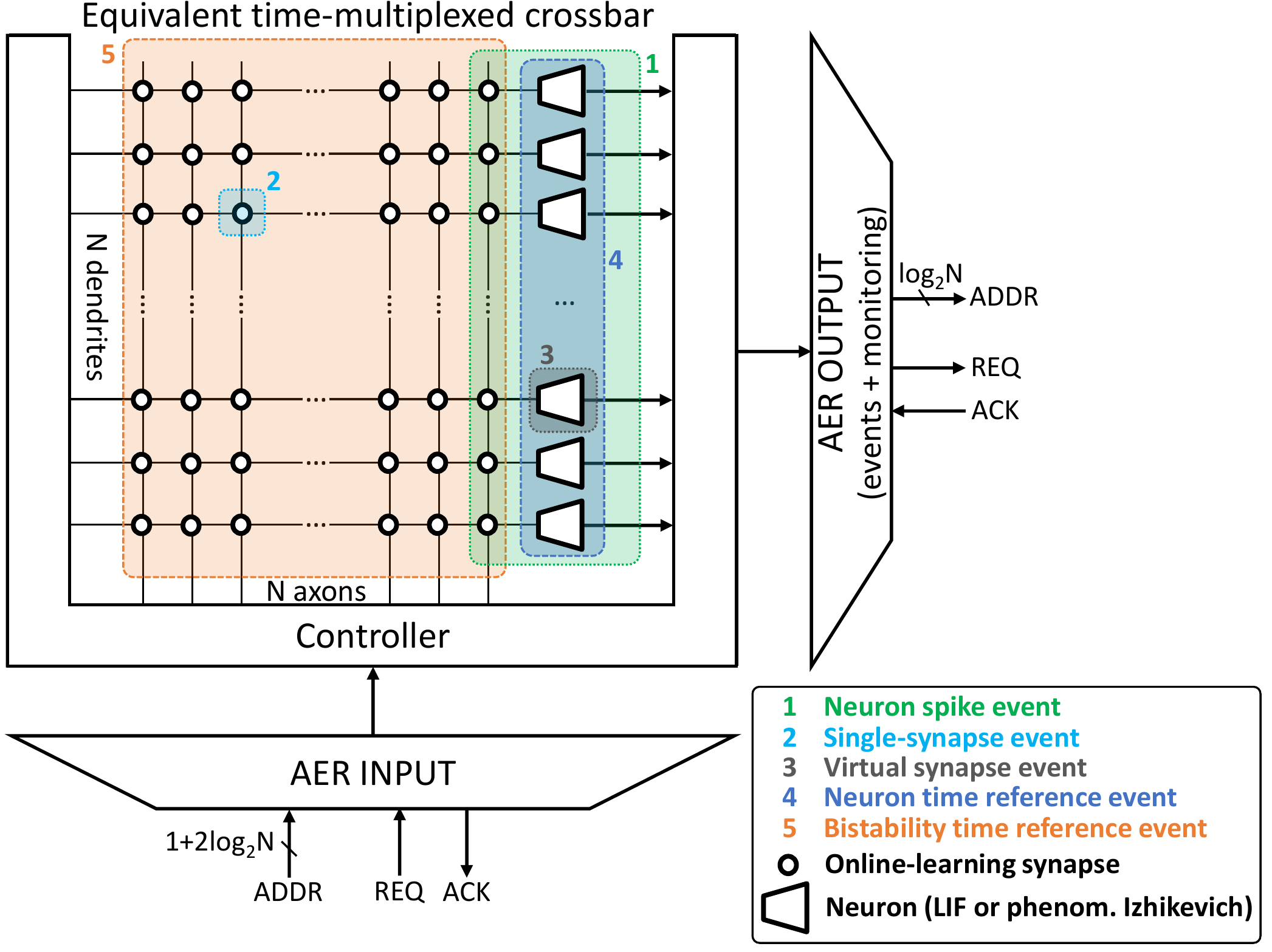}
\vspace*{-2mm}
\caption{Equivalent N-neuron N$^2$-synapse time-multiplexed crossbar architecture. Input AER addresses cover neuron, single-synapse, virtual synapse and time reference events for neurons and synaptic bistability.\vspace*{-2mm}}
\label{fig_crossbar}
\end{figure}

The crossbar architecture implemented in the ODIN SNN is shown in Fig.~\ref{fig_crossbar}. In order to minimize the silicon area by leveraging the high-speed operation of digital logic, a time multiplexing approach is followed in which neurons and synapses are updated in a sequential manner, not in parallel. The individual states of N neurons and N$^2$ synapses are stored in on-chip SRAMs, a controller handles the neuron and synapse updates to emulate a \mbox{N$\times$N} crossbar, where each of the N neurons thus has a fan-in of N online-learning synapses.

Input and output events at the chip level are handled through address-event representation (AER) buses, a popular four-phase handshake communication scheme, which is the \textit{de facto} standard for SNNs~\cite{Mortara94,Boahen00}. As ODIN is implemented using a synchronous digital implementation flow (Section~\ref{sec_mesu}), in order for the AER to be used in an asynchronous fashion, a double-latching synchronization barrier has been placed on the \texttt{REQ} and \texttt{ACK} lines of the input and output AER interfaces, respectively, to limit metastability issues. In order to increase versatility in experimentation and testing of the ODIN SNN, the input AER bus is extended in order to represent a wide variety of event types.
\begin{enumerate}
\item Neuron spike event -- It is the standard operation of ODIN, which can be triggered externally from the input AER interface or internally from the embedded neurons. If the log$_2$N-bit address of a source neuron $i$ is received over AER, all N neurons of ODIN will be updated with the synaptic weight found at synapse $i$ in their respective N-synapse dendritic tree: each neuron spike event thus leads to N SOPs. An SDSP weight update will also be applied to these synapses~(Section~\ref{ssec_synapse}).
\item Single-synapse event -- Two log$_2$N-bit addresses are provided in the \texttt{ADDR} field of the input AER interface: the address of a source neuron $i$ and the address of a destination neuron $j$. This event is handled similarly to an AER neuron spike event, but only the neuron $j$ of ODIN will be updated, together with a single SDSP weight update to synapse $i$ of neuron~$j$.
\item Virtual synapse event -- The log$_2$N-bit address of a target neuron $j$ of ODIN and a fixed weight value are provided in the \texttt{ADDR} field, no synapse of ODIN is read nor updated.
\item Neuron time reference event -- A specific address is used to provide an external time reference event to all neurons, which defines the time constant of the Izhikevich behaviors of the phenomenological neuron model~(Section~\ref{ssec_neuron}).
\item Bistability time reference event -- A specific address is used to trigger the bistability mechanism of all synapses~(Section~\ref{ssec_synapse}).
\end{enumerate}

Therefore, in order to represent these five event types with all neuron, synapse and specific time reference control addresses, the input AER bus needs an \texttt{ADDR} field width of 1+2log$_2$N bits~(Fig.~\ref{fig_crossbar}).

Two operation modes can be chosen for the log$_2$N-bit output AER bus. The first one is the standard operation mode in which, as soon as one of the N neurons emits an output spike event, the source neuron address is transmitted on the log$_2$N-bit \texttt{ADDR} field of the AER output. The second one is a non-standard operation mode for monitoring purposes of a specific neuron or synapse, AER events containing state data are generated. The target neuron and synapse addresses to be monitored are configurable.

\begin{figure}[!t]
\centering
\vspace*{1.5mm}
\noindent\includegraphics[width=1.00\columnwidth]{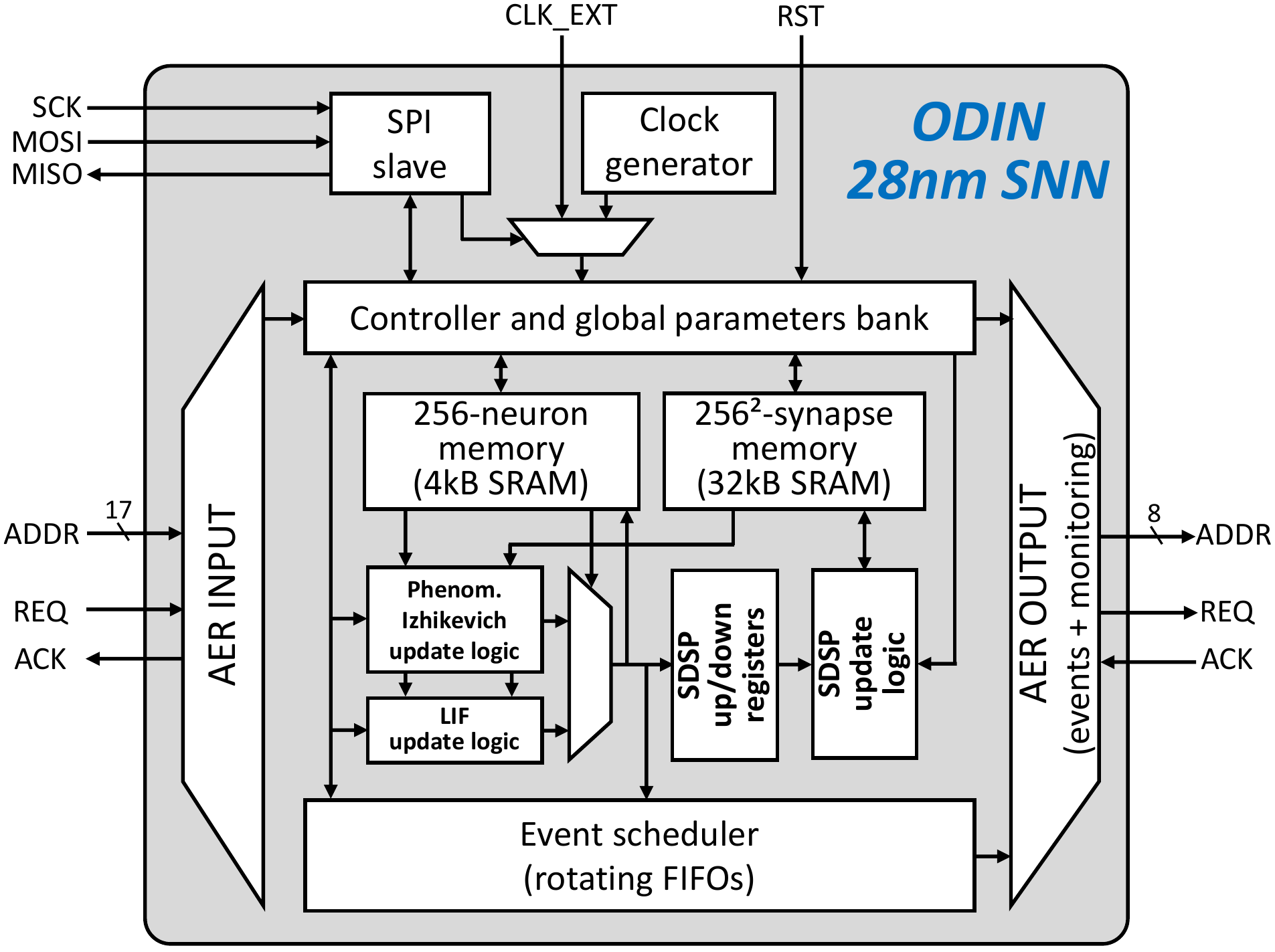}
\caption{Block diagram of the ODIN online-learning digital spiking neuromorphic processor, which implements the architecture shown in Fig.~\ref{fig_crossbar} for N$=$256 neurons.}
\vspace*{1.5mm}
\label{fig_arch}
\end{figure}

\begin{figure}[!t]
\centering
\vspace*{1.5mm}
\noindent\includegraphics[width=1.0\columnwidth]{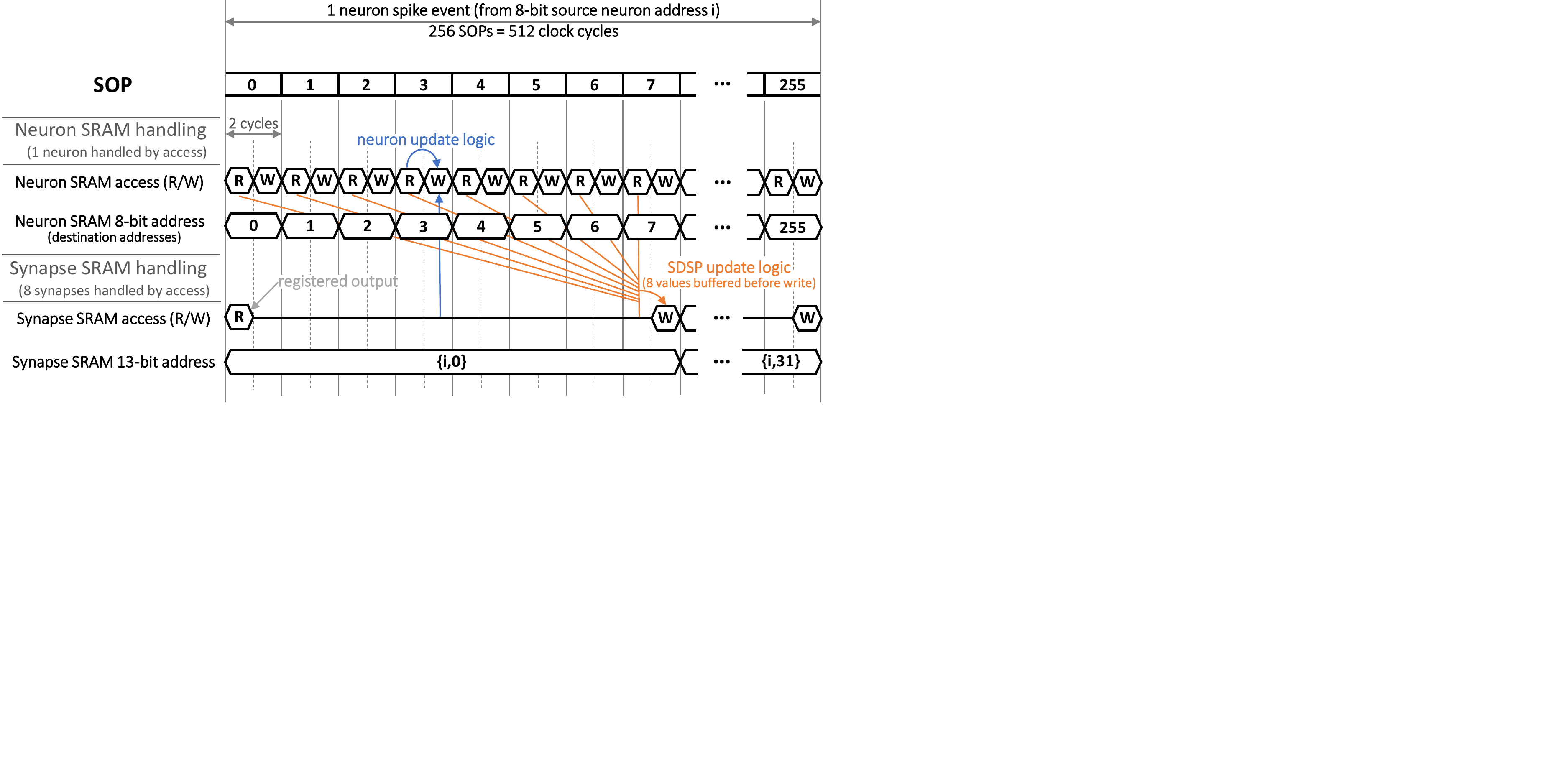}
\caption{Timing diagram of ODIN for a neuron spike event from 8-bit source neuron address $i$, leading to 256 SOPs. Each SOP lasts two clock cycles. First,  the current state of a destination neuron with 8-bit address $j$ is retrieved. Second, the current state of neuron $j$ and the associated synaptic weight corresponding to the source neuron $i$ are provided to the neuron update logic, the updated state of neuron $j$ is then written back to the neuron memory. As ODIN has 4-bit synapses and the chosen synapse SRAM has 32-bit words for density purposes, read and write operations to the synapse memory handle 8 synapses at a time in order to minimize memory accesses. The SDSP update logic takes information from the Calcium variable and membrane potential of the destination neurons in order to update the associated synaptic weights.}
\vspace*{1.5mm}
\label{fig_clock}
\end{figure}

Fig.~\ref{fig_arch} gives a block diagram overview of the ODIN SNN, which implements the crossbar architecture presented in Fig.~\ref{fig_crossbar} for N$=$256 neurons. The individual states and parameters of the 256 neurons and 64k synapses are respectively stored in 4kB and 32kB high-density single-port foundry SRAMs while a global controller manages time multiplexing of the synapse and neuron update logic~(Sections~\ref{ssec_synapse} and~\ref{ssec_neuron}), the associated timing diagram is shown in Fig.~\ref{fig_clock}. All SNN-level control registers as well as individual neuron and synapse states and parameters can be read and written through a conventional SPI bus, while input and output events at the chip level are handled through AER buses. Finally, a scheduler based on rotating FIFOs~(Section~\ref{ssec_scheduler}) arbitrates and orders external and internally-generated events, including bursts of spikes induced by Izhikevich bursting behaviors.

\subsection{Synapse array} \label{ssec_synapse}

\begin{figure}[!t]
\centering
\noindent\includegraphics[width=0.97\columnwidth]{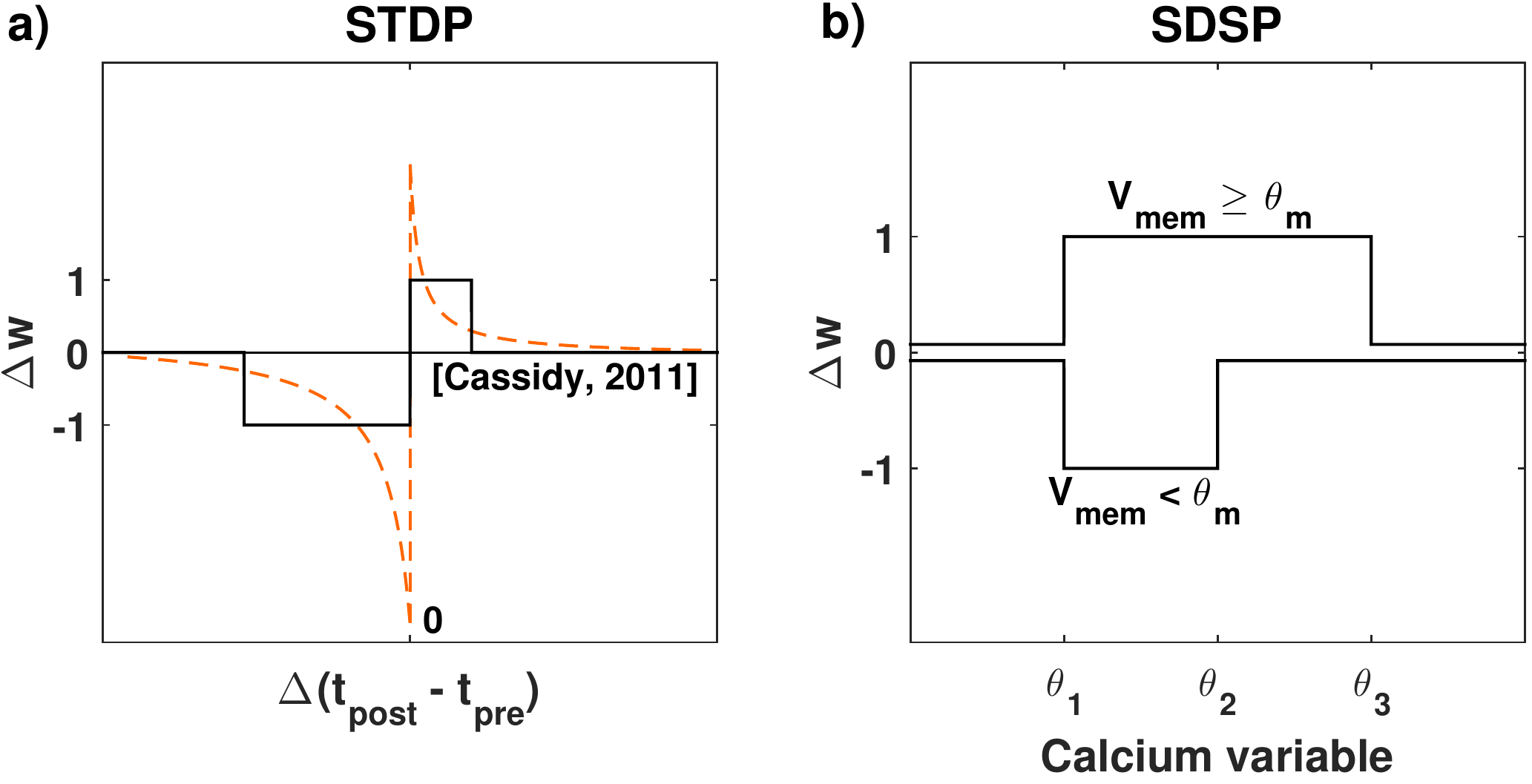}
\caption{(a) STDP learning rule with digital approximation proposed by Cassidy \textit{et al.}~\cite{Cassidy11}. (b) SDSP learning rule proposed by Brader \textit{et al.}~\cite{Brader07}. Adapted from~\cite{Frenkel17a}.\vspace*{2mm}}
\label{fig_STDP_SDSP}
\end{figure}

In order to embed online learning in each synapse of the 256$\times$256 crossbar array, an efficient digital implementation of a learning rule is required. Representations of the STDP learning rule with the digital approximation proposed by Cassidy~\textit{et~al.}~\cite{Cassidy11} and of the SDSP learning rule proposed by Brader~\textit{et~al.}~\cite{Brader07} are illustrated in Figs.~\ref{fig_STDP_SDSP}a and~\ref{fig_STDP_SDSP}b, respectively. While the STDP learning rule relies on the relative pre- and post-synaptic spike times t$_\text{pre}$ and t$_\text{post}$, the SDSP learning rule induces an update each time a pre-synaptic spike occurs. The conditions of SDSP for synaptic weight increment (i.e. $\Delta w=+1$) and decrement (i.e. $\Delta w=-1$) follow Eq.~(\ref{eq_sdsp}) and depend only on the state of the post-synaptic neuron at the time of the pre-synaptic spike, i.e. the value of the membrane potential $V_{mem}$ (compared to threshold $\theta_m$) and of the Calcium concentration Ca (compared to thresholds $\theta_1$, $\theta_2$ and $\theta_3$), where the Calcium concentration represents an image of the post-synaptic firing activity~(Section~\ref{ssec_neuron}). Therefore, all necessary SDSP computations are offloaded to neurons and do not need to be replicated inside each synapse, leading to substantial resource and area savings compared to STDP. Performance of STDP and SDSP learning rules is similar, while thresholds on Calcium concentration in SDSP result in an embedded overfitting prevention mechanism, referred to as stop-learning conditions by Brader \textit{et al.}~\cite{Brader07}. Detailed explanations and quantitative results comparing digital STDP and SDSP implementations can be found in~\cite{Frenkel17a}.%
\begin{equation}\label{eq_sdsp}
\hspace*{-4mm}\begin{cases}
w \rightarrow w + 1  ~~ \text{if~~} V_{\text{mem}}(t_{\text{pre}})\geq\theta_{m}, ~\theta_1\leq \text{Ca}(t_{\text{pre}})<\theta_3 \\
w \rightarrow w - 1  ~~ \text{if~~} V_{\text{mem}}(t_{\text{pre}})<\theta_{m}, ~\theta_1\leq \text{Ca}(t_{\text{pre}})<\theta_2
\end{cases}
\end{equation}

\begin{figure}[!t]
\centering
\noindent\includegraphics[width=0.97\columnwidth]{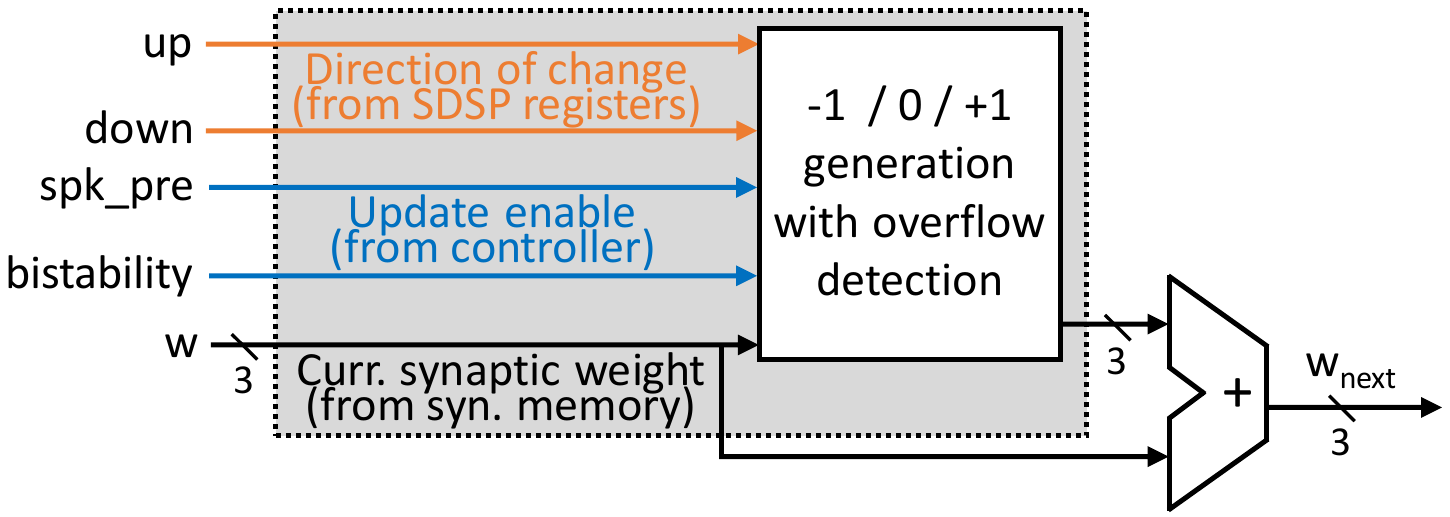}
\caption{SDSP update logic for time-multiplexed implementation of the standalone SDSP synapse proposed in~\cite{Frenkel17a}. The different blocks interacting with the SDSP update logic are shown in Fig.~\ref{fig_arch}.}
\label{fig_synapseArch}
\end{figure}

\begin{figure*}[!t]
\centering
\vspace*{-1.2mm}
\noindent\includegraphics[width=0.88\textwidth]{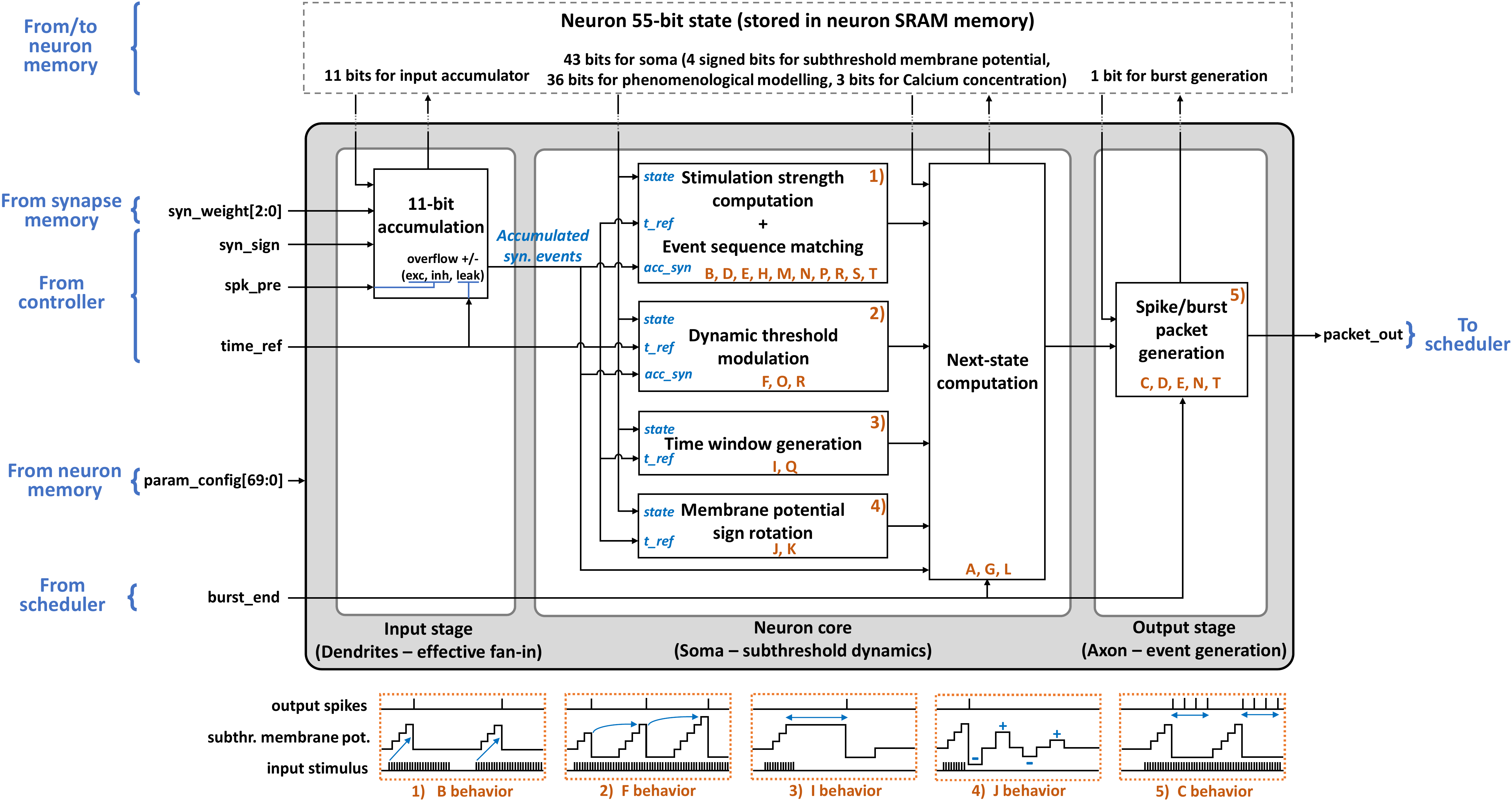}
\caption{Three-stage architecture of the proposed time-multiplexed phenomenological digital neuron update logic, extended from~\cite{Frenkel17b}. Letters inside the different combinational logic blocks indicate the Izhikevich behaviors that are phenomenologically generated, following the numbering convention provided in Fig.~\ref{fig_20behav}. A sample behavior illustration is also provided for each combinational logic block to illustrate their functionality on a practical case, as outlined in blue in the behavior illustrations. The different blocks interacting with the neuron update logic are shown in Fig.~\ref{fig_arch}.\vspace*{-2.3mm}}
\label{fig_neuronArch}
\end{figure*}

Each synapse of ODIN consists of a weight and 1 bit of mapping table to enable or disable online learning locally. A 3-bit resolution was chosen for the weight in order to obtain a resolution similar to the one presented by Brader~\textit{et~al.} in~\cite{Brader07}, which results in a 4-bit synapse including the weight and the mapping table bit. All thresholds $\theta_m$, $\theta_1$, $\theta_2$ and $\theta_3$ of each neuron can be individually programmed through SPI~(Section~\ref{ssec_neuron}). We implemented a time-multiplexed version of the digital standalone SDSP synapse that we previously proposed in~\cite{Frenkel17a}~(Fig.~\ref{fig_synapseArch}). The \texttt{up} and \texttt{down} signals, coming from the SDSP registers shown in Fig.~\ref{fig_arch}, represent the increment/decrement conditions of SDSP updates in Eq.~(\ref{eq_sdsp}). The controller provides control signals conditionally enabling the updates: \texttt{spk\_pre} is linked to incoming pre-synaptic spikes and \texttt{bistability} is associated to bistability time reference events received externally through specific AER events. The bistability mechanism is optional: each time a bistability time reference event is received, if the current synaptic weight is above (resp. below) half its dynamic, it increments (resp. decrements) toward a high (resp. low) state. Bistability prevents undesired fluctuations in the case of spontaneous network activity and supports the SDSP stochastic selection mechanism for long-term potentiation (LTP) and long-term depression (LTD) with Poisson-distributed network activities~\cite{Brader07,Frenkel17a}. Each synaptic operation (SOP) in ODIN takes two clock cycles, as shown in Fig.~\ref{fig_clock}.

The time-multiplexed SDSP learning rule is compatible with standard foundry SRAMs in order to store the synaptic states, as opposed to the digital time-multiplexed probabilistic STDP synapse proposed by Seo~\textit{et al.} in~\cite{Seo11}, which requires a custom dual-port SRAM with row and column accesses. Beyond design time savings, single-port foundry SRAMs benefit from pushed rules for foundry bitcells and avoid the area overhead of custom SRAM designs, which use DRC rules for logic~\cite{Chen11}. This efficient time-multiplexed digital SDSP learning rule is implemented at high density: each 4-bit synapse of ODIN embeds online learning and occupies only 0.68$\mu$m$^2$, taking into account the synapse SRAM area, the SDSP update logic and the addition of a Calcium variable into the neurons.

\subsection{Neuron array} \label{ssec_neuron}

In order to leverage spike-based processing, time must be included into computation~\cite{Arthur06}. As the standard LIF neuron model only behaves like an integrator, it lacks the ability and behavior repertoire to efficiently compute using temporal information~\cite{Indiveri10}. Therefore, beyond the rate code which is popular due to its simplicity but has an inefficient spike use, behavior versatility is crucial to explore other codes that encode higher amounts of data bits per spike by leveraging time, such as the timing code or the synchrony-based code~\cite{Thorpe01,Yu13}. The 20 Izhikevich behaviors of biological spiking neurons~\cite{Izhikevich04}, offer a variety of ways to capture time into computation. For example, phasic spiking captures the stimulation onset~\cite{Izhikevich04} and could be useful for the synchrony-based code in which neurons are allowed to emit a single spike~\cite{Thorpe01}. Spike frequency adaptation is useful to encode time since stimulation onset~\cite{Izhikevich04}, while both spike frequency adaptation and threshold variability can be used to implement forms of homeostatic plasticity, which allows stabilizing the global network activity~\cite{Davies18,Qiao17}. Spike latency can emulate axonal delays, which are useful to induce temporal dynamics in SNNs and to enhance neural synchrony~\cite{Sheik12}, while resonant behaviors allow selectively responding to specific frequencies and spike time intervals, thus enabling the timing code.

ODIN comprises two neuron models~(Fig.~\ref{fig_arch}): each of the 256 neurons of ODIN can be individually chosen between a standard 8-bit LIF neuron model and a custom phenomenological model of the 20 Izhikevich behaviors. As the neurons are time-multiplexed and their states and parameters are stored in a standard single-port SRAM, the use of the combinational update logic of the LIF model or of the phenomenological model is determined by a neuron parameter bit. Both neuron models are extended with a 3-bit Calcium variable and associated Calcium threshold parameters to enable synaptic SDSP online learning~(Section~\ref{ssec_synapse}). As soon as one neuron spikes, the value of its Calcium variable is incremented. The Calcium leakage time constant is configurable and depends on the neuron time reference events provided externally through specific AER events. The Calcium variable of a neuron thus represents an image of its firing activity.

The three-stage architecture of the proposed phenomenological neuron is shown in Fig.~\ref{fig_neuronArch}, it is the time-multiplexed and Calcium-extended version of the standalone neuron we previously proposed in~\cite{Frenkel17b}. The characteristics and behaviors of each neuron are independently controlled with a 70-bit \texttt{param\_config} parameter array stored in the neuron memory, next to the 55-bit neuron state. Each neuron of ODIN thus requires 126 parameter and state bits, including one bit to select between LIF and phenomenological neuron models. These parameters and the initial state of all neurons are preloaded during the initial SPI configuration of ODIN. The time constants of the neuronal dynamics range from biological- to accelerated-time depending on the frequency of external time reference events. The neuron is entirely event-driven, the state is updated only upon reception of synaptic or time reference events, no mathematical equation needs to be solved at each integration timestep. While piecewise linear approximations have also been proposed to save the overhead of coupled non-linear mathematical equation solvers (e.g., \cite{Soleimani12,Yang15}), they still require to update all neuron states after each integration timestep, which induces high computational overhead, especially at accelerated time constants. Therefore, the proposed neuron model implements the behaviors and main functionalities of the Izhikevich neuron model~at low area and computational cost, not its detailed dynamics.

The input stage can be assimilated to dendrites. The effective fan-in of the neuron is controlled by an 11-bit accumulator with configurable depth. When an input event occurs, the controller asserts the \texttt{spk\_pre} signal while the 3-bit synaptic weight value associated to the pre-synaptic neuron is retrieved from the synapse memory~(Fig.~\ref{fig_arch}). The sign of the contribution is contained in the \texttt{syn\_sign} signal depending on the pre-synaptic neuron address: each source neuron address can be defined as excitatory or inhibitory in the global parameters of ODIN, which allows easy implementations of neuromorphic computational primitives with shared or global inhibitory neurons, such as winner-take-all (WTA) networks~\cite{Neftci13}. When a time reference \texttt{time\_ref} event occurs, leakage with configurable strength is applied to the input accumulator.

The neuron core can be assimilated to the soma. Following the \textit{spike} or \textit{pulse} model assumption for cortical neurons, information lies in spike timings, not in spike shapes~\cite{Koch99}. Therefore, only subthreshold dynamics are modeled in the neuron core in order to dedicate resolution bits to the informative part of the neuron membrane potential. Subthreshold dynamics are driven by four blocks that capture the functionalities necessary to phenomenologically describe the 20 Izhikevich behaviors, as indicated in Fig.~\ref{fig_neuronArch} by letters corresponding to the behavior numbering of Fig.~\ref{fig_20behav}. These blocks are stimulated by accumulated synaptic events resulting from positive and negative overflows of the input stage accumulator, while all time-dependent computations are triggered by time reference events. The first block computes the strength and duration of the ongoing input stimulation to assess if specific stimulation sequences are matched, which captures phasic, mixed, class 2, rebound, bistability, accommodation and inhibition-induced behaviors. The second block allows the neuron firing threshold to depend on the input stimulation or on the output firing patterns, which captures threshold variability and spike frequency adaptation behaviors. The third block of time window generation allows delaying some neuron operations like firing and membrane potential resetting, which captures spike latency and depolarizing after-potential behaviors. Beyond emulating axonal delays with the spike latency behavior, refractory delays are also implemented with a configurable refractory period. The fourth block allows rotating the sign of the subthreshold membrane potential in order to capture the subthreshold oscillation and resonator behaviors. The activation, extent and effects of all four blocks used in phenomenological modeling are configured using the neuron parameters in \texttt{param\_config}.

Finally, the output stage can be assimilated to the axon. When firing conditions are met in the neuron core, the output stage generates a neuron output event packet~\texttt{packet\_out} containing the number of spikes and inter-spike interval (ISI) to be generated, following parameters in \texttt{param\_config}. As the neuron state is updated only after input events and as bursts usually have an ISI shorter than neuron time reference events~(Fig.~\ref{fig_20behav}), the neuron alone is not able to update itself in order to generate bursts. Therefore, neuron event packets are sent to the scheduler of ODIN~(Section~\ref{ssec_scheduler}) which arbitrates the generation of spikes and bursts from all neurons. When a neuron emits a burst event, its membrane potential is reset and then locked down until the scheduler signals the end of burst generation using the \texttt{burst\_end} signal.

\subsection{Scheduler} \label{ssec_scheduler}

\begin{figure}[!t]
\centering
\noindent\includegraphics[width=0.96\columnwidth]{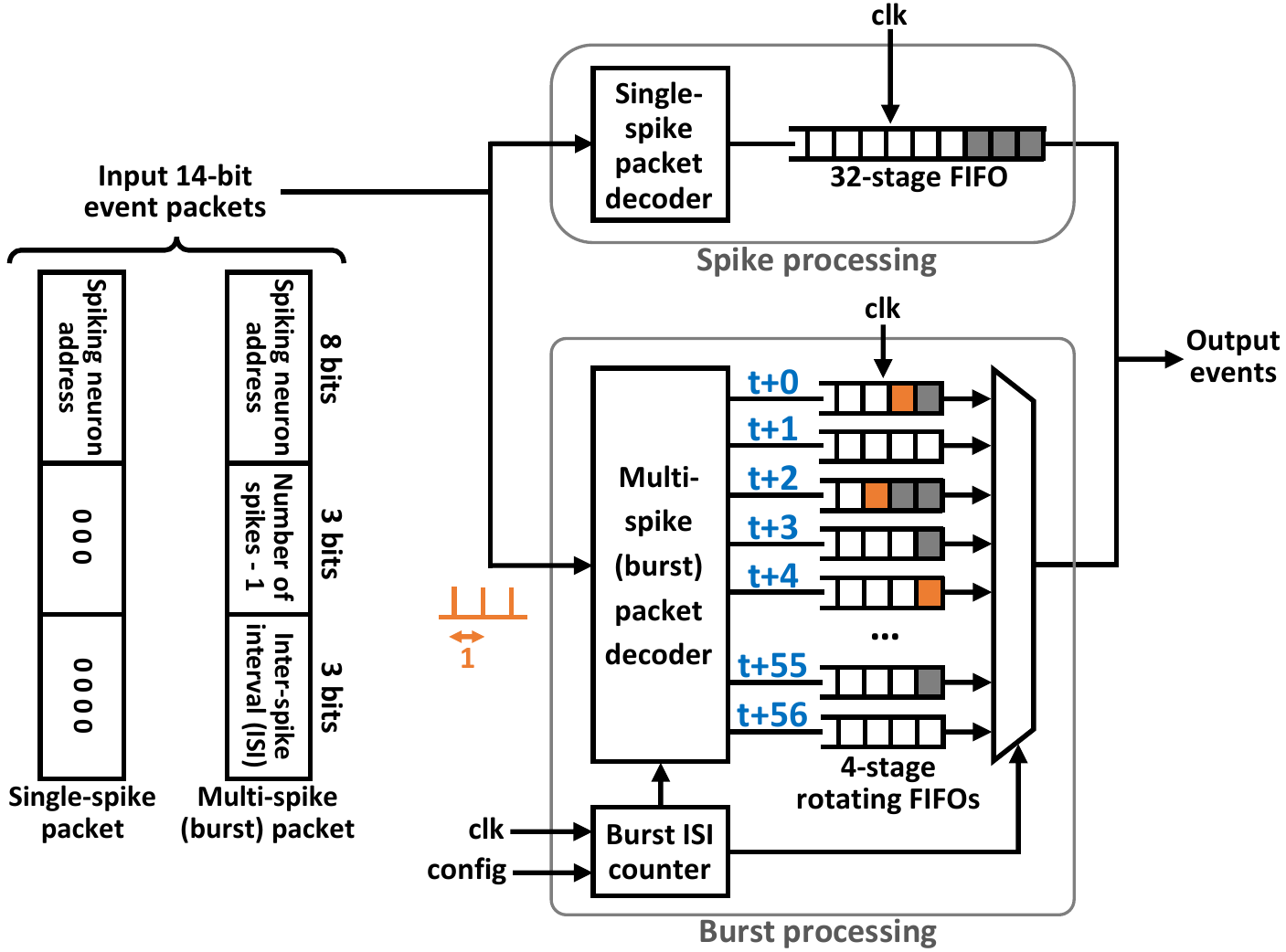}
\caption{Scheduler architecture and contents of event packets sent by spiking neurons. Handling bursts issued by the Izhikevich behaviors requires the use of rotating FIFOs in order to reproduce the inter-spike interval (ISI) value that is specified in the incoming burst packet. Orange spikes illustrate a 3-spike \mbox{1-timestep-ISI} burst example being stored in the scheduler rotating FIFOs, where grayed-out stages represent previously-filled spike events waiting to be processed by the controller.}
\label{fig_schedulerArch}
\end{figure}

The scheduler is shown in Fig.~\ref{fig_schedulerArch} and can be assimilated to a priority-based ordered FIFO~\cite{Moon00}. Its purpose is to handle spiking and bursting events from all neurons and to arbitrate between external and internally-generated neuron events. Spiking and bursting neurons of ODIN send 14-bit event packets to the scheduler. The packets contain the 8-bit address of the source neuron, a 3-bit field indicating the number of spikes to be generated minus one and a 3-bit field quantifying the ISI. The two latter parameters are zero in the case of single-spike events. All single-spike events are stored in a FIFO of 32$\times$ 8-bit stages, where each stage stores the address of the source spiking neuron. In the global priority-based ordered FIFO structure of the scheduler, this single-spike FIFO always has the highest priority: all single-spike events are handled as soon as they are available.

All multi-spike events (i.e. burst events) are separated by the burst packet decoder into multiple single-spike events and split among rotating 4-stage FIFOs, corresponding to successive timesteps. All FIFOs are assigned a priority token inspired from the FIFO priority queue architecture proposed in~\cite{Moon00}. In order to generate the ISI of the different bursts, the scheduler generates its own timestep locally in the burst ISI counter, which is configurable to range from biological to accelerated time constants. As soon as one timestep is elapsed, the priorities of all 4-stage FIFOs are rotated. As neuron event packets can encode bursts of up to 8 spikes with an ISI of up to 7 timesteps, a total of (7$\times$8)+1=57 rotating 4-stage FIFOs are necessary to encode all priorities, which are denoted in Fig.~\ref{fig_schedulerArch} by reference to the burst ISI timestep $t$, where the highest-priority current timestep is $t+0$. To illustrate the scheduler operation, we show in Fig.~\ref{fig_schedulerArch} an input neuron event packet describing a 3-spike \mbox{1-timestep-ISI} burst as an example. Three single-spike events will be generated by the burst packet decoder toward FIFOs associated with local ISI timesteps $+0$, $+2$ and $+4$. The FIFO stages have a 9-bit width, they contain the 8-bit address of the source spiking neuron with one bit marking the last spike of the burst, currently stored in the timestep $+4$ FIFO. The spike at timestep $+0$ will be processed immediately by the controller. After two timestep ticks of the ISI counter and two rotations of FIFO priorities, the second spike of the burst gets to timestep $+0$ and is processed by the controller. After two other ISI counter ticks, the last spike of the burst is retrieved by the controller: as one bit indicates that this spike ends a burst, the scheduler unlocks state updates of the associated source neuron~(Section~\ref{ssec_neuron}).%

\section{Specifications, Measurements and Benchmarking Results} \label{sec_mesu}

The ODIN neuromorphic processor was fabricated in ST Microelectronics 28-nm fully-depleted silicon-on-insulator (FDSOI) CMOS process, with a mix of 8-track LVT libraries with upsized gate length in order to reduce leakage (poly-biasing technique by 4nm and 16nm). ODIN is implemented following a standard fully-synthesizable synchronous digital implementation flow. Fig.~\ref{fig_chipSpec}a shows the chip microphotograph. ODIN occupies an area of only 320$\mu$m$\times$270$\mu$m post-shrink, the remaining chip area is used by unrelated blocks. Fig.~\ref{fig_chipSpec}a also shows a zoom on the floorplan view, highlighting that the synapse and neuron single-port foundry SRAMs occupy 50\% and 15\% of the area, respectively. These numbers directly provide the area occupied by the 64k synapses and 256 neurons of ODIN as the associated combinational update logic area is negligible compared to the SRAM sizes. The scheduler, controller, neuron and synapse update logic, SPI and AER buses are implemented with digital standard cells in the remaining area of the floorplan, 80\% of which is occupied by the scheduler. All measurement results presented in this section were accurately predicted by a Python-based simulator we developed for ODIN: one-to-one hardware/software correspondence is thus ensured.

\subsection{Chip specifications and measurements} \label{ssec_chip_spef}

\begin{figure}[!t]
\centering
\noindent\includegraphics[width=0.967\columnwidth]{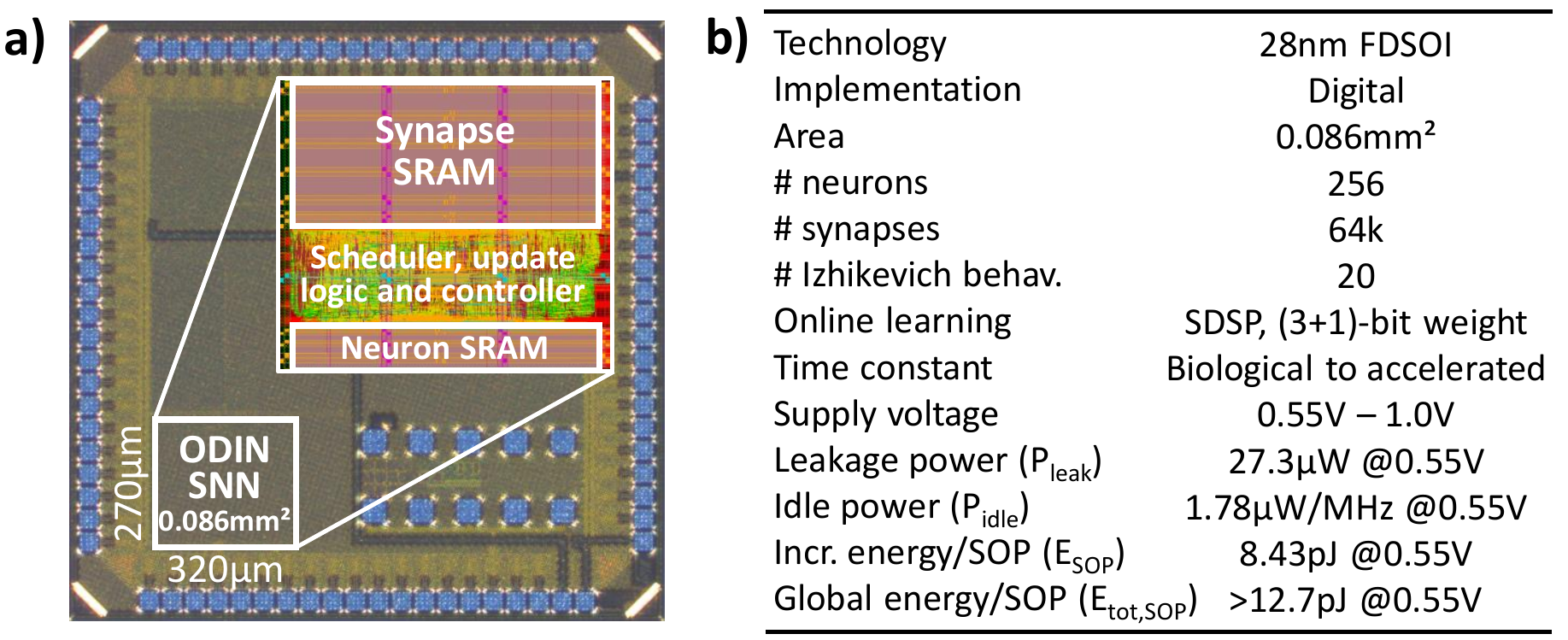}
\caption{(a) Chip microphotograph with zoom on floorplan view. Chip area outside the 320$\mu$m$\times$270$\mu$m of the ODIN neuromorphic processor is occupied by unrelated blocks. (b) Specifications and measurements. $P_\text{leak}$, $P_\text{idle}$ and $E_\text{SOP}$ are parameters of the ODIN power consumption model in Eq.~(\ref{eq_power}).}
\label{fig_chipSpec}
\end{figure}

Specifications of the 64k-synapse 256-neuron ODIN spiking neuromorphic processor are summarized in Fig.~\ref{fig_chipSpec}b. It embeds SDSP-based online learning with each synapse consisting of a 3-bit weight and one bit of mapping table, while phenomenological neurons are able to exhibit all 20 Izhikevich behaviors from biological to accelerated time constants. A total of 9 ODIN chips were available for tests, the power and energy measurements reported in Fig.~\ref{fig_chipSpec}b are average values across ODIN test chips and have been acquired at a temperature of 24$^\circ$C. ODIN remains fully functional while scaling the supply voltage down to 0.55V, where the maximum operating frequency of the chip is 75MHz, against 100MHz at the nominal supply voltage of 0.8V. The different contributions to the power consumption $P$ of ODIN are summarized in Eq.~(\ref{eq_power}):
\begin{equation}\label{eq_power}
P = P_\text{leak} + P_\text{idle}\times f_\text{clk} + E_\text{SOP}\times r_\text{SOP},
\end{equation}
where $P_\text{leak}$ is the leakage power without clock activity and $P_\text{leak} + P_\text{idle}\times f_\text{clk}$ is the power consumption of ODIN without any activity in the network when a clock of frequency $f_\text{clk}$ is connected. $E_\text{SOP}$ is the energy paid for each SOP, which includes the contributions of reading and updating the synaptic weight according to the SDSP learning rule, reading and updating the associated neuron state, as well as the controller and scheduler overheads. Finally, $r_\text{SOP}$ is the average SOP processing rate, the maximum SOP rate ODIN can handle is equal to $f_\text{clk}/2$ as each SOP takes two clock cycles to complete~(Fig.~\ref{fig_clock}).

At 0.55V, $P_\text{leak}$ is $27.3\mu W$ and $P_\text{idle}$ is $1.78\mu W/\text{MHz}$. In order to determine $E_\text{SOP}$, we saturate the scheduler with neuron spike event requests, each consisting in 256 SOPs~(Section~\ref{sec_arch}), so that the achieved SOP rate $r_\text{SOP}$ is at its maximum with one SOP processed every two clock cycles. By measuring the chip power consumption and subtracting leakage and idle power, we obtain an energy per SOP $E_\text{SOP}$ of 8.43pJ.

\begin{figure}[!t]
\centering
\noindent\includegraphics[width=0.9\columnwidth]{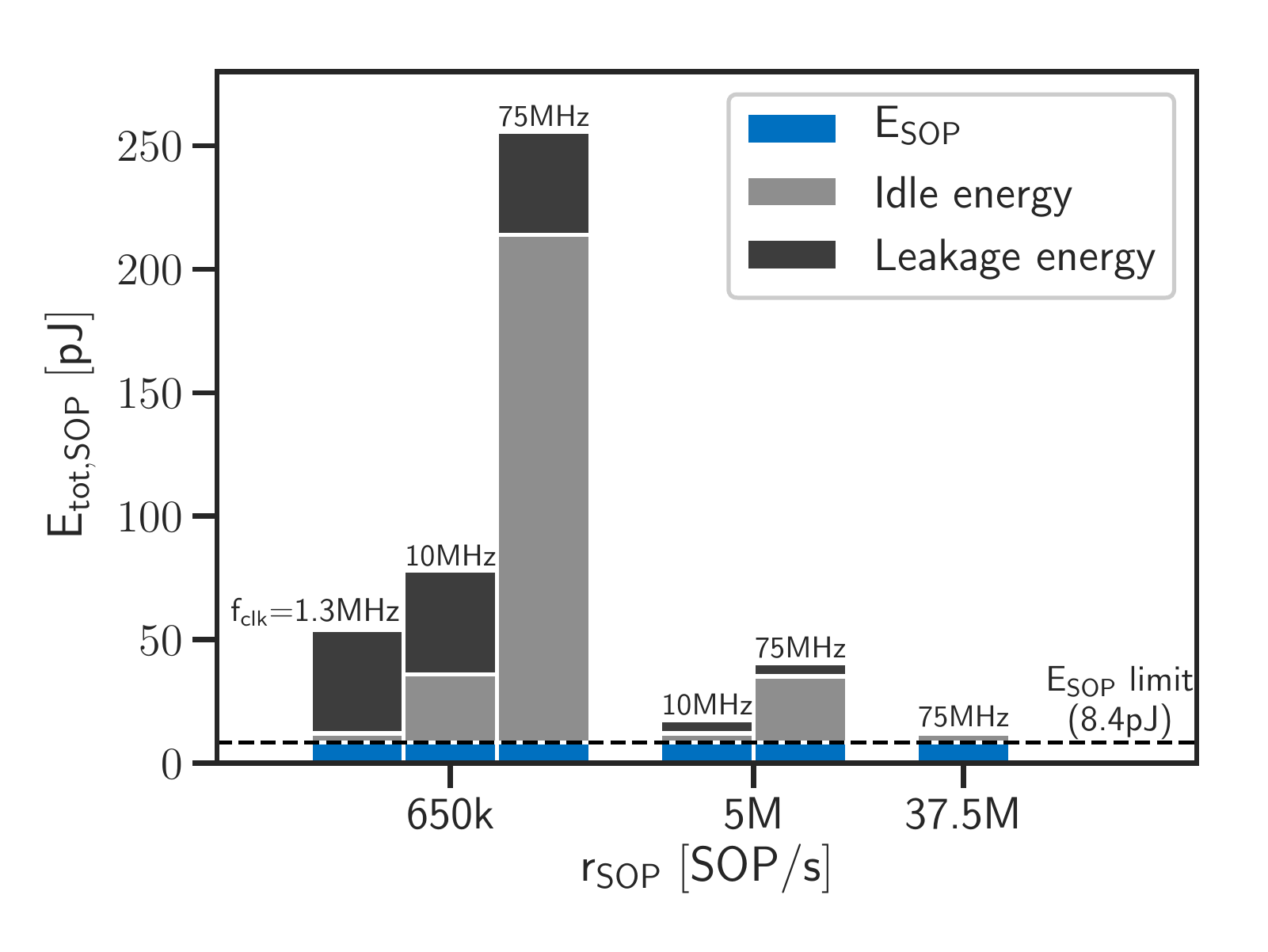}
\vspace*{-2.5mm}
\caption{Energy breakdown of the global SOP energy $E_\text{tot,SOP}$ at 0.55V, illustrating the contributions from the incremental SOP energy $E_\text{SOP}$, leakage energy and idle energy for different choices of $r_\text{SOP}$ and $f_\text{clk}$. The energy efficiency of ODIN is the highest in accelerated time as it approaches the incremental energy $E_\text{SOP}$.\vspace*{-2.6mm}}
\label{fig_power_model}
\end{figure}

Another commonly-employed definition for the energy per SOP can be found in Eq.~(\ref{eq_powerTot}), where the whole chip power consumption $P$ is divided by the SOP rate $r_\text{SOP}$, without subtracting contributions from leakage and idle power (see Section~\ref{sec_disc} and Table~\ref{table_SoA} for an energy per SOP summary across state-of-the-art neuromorphic chips). In order to avoid confusion in the subsequent text, we will denote $E_\text{SOP}$ as the \textit{incremental} energy per SOP and $E_\text{tot,SOP}$ as the \textit{global} energy per SOP. The former is appropriate for the definition of the power model in Eq.~(\ref{eq_power}) and is useful to accurately predict the power consumption of ODIN based on its activity and clock frequency, while the latter is more application-driven and representative of the real energy per SOP performance.
\begin{equation}\label{eq_powerTot}
E_\text{tot,SOP} = P/r_\text{SOP}
\end{equation}
At a clock frequency of 75MHz, the maximum $r_\text{SOP}$ is equal to 37.5MSOP/s and the measured power consumption $P$ of ODIN is 477$\mu$W at 0.55V. At this accelerated-time rate and maximum $r_\text{SOP}$, the global energy per SOP $E_\text{tot,SOP}$ is equal to  12.7pJ, which is dominated by dynamic power as the influence of leakage is negligible (6\% of the total power) and the idle power accounts for 28\% of the total power. In order to estimate $E_\text{tot,SOP}$ when ODIN operates at a biological time constant, an order of magnitude can be found by assuming that all 256 neurons of ODIN spike at a rate of 10Hz. Each neuron event leading to 256 SOPs, the corresponding SOP rate $r_\text{SOP}$ is equal to 650kSOP/s, a clock frequency of at least 1.3MHz is thus sufficient to operate ODIN at a biological time constant. In this biological-time regime, the measured power consumption $P$ of ODIN is 35$\mu$W at 0.55V. Therefore, the global energy per SOP $E_\text{tot,SOP}$ at biological time is equal to 54pJ, which is dominated by leakage (78\% of the total power). These results highlight that ODIN achieves a better efficiency in accelerated-time operation thanks to leakage amortization over more SOPs per second, as shown in Fig.~\ref{fig_power_model}. In this regime, $E_\text{tot,SOP}$ approaches the incremental SOP energy $E_\text{SOP}$. In order to minimize the power consumption, the clock frequency of ODIN needs to be set according to the target application, taking into account the expected network activity and the required temporal resolution, which vanishes when $r_\text{SOP}$ approaches its maximum value of $f_\text{clk}/2$. The power model of Eq.~(\ref{eq_power}) allows evaluating this tradeoff by computing the optimal operating point of ODIN and the resulting power consumption. As the power model of Eq.~(\ref{eq_power}) was derived at a room temperature of 24$^\circ$C and 0.55V, care should be taken with the operating conditions of ODIN in the target application (e.g.,~the contribution of leakage increases in high-temperature operating conditions).

\begin{figure*}[!t]
\centering
\noindent\includegraphics[width=0.9815\textwidth]{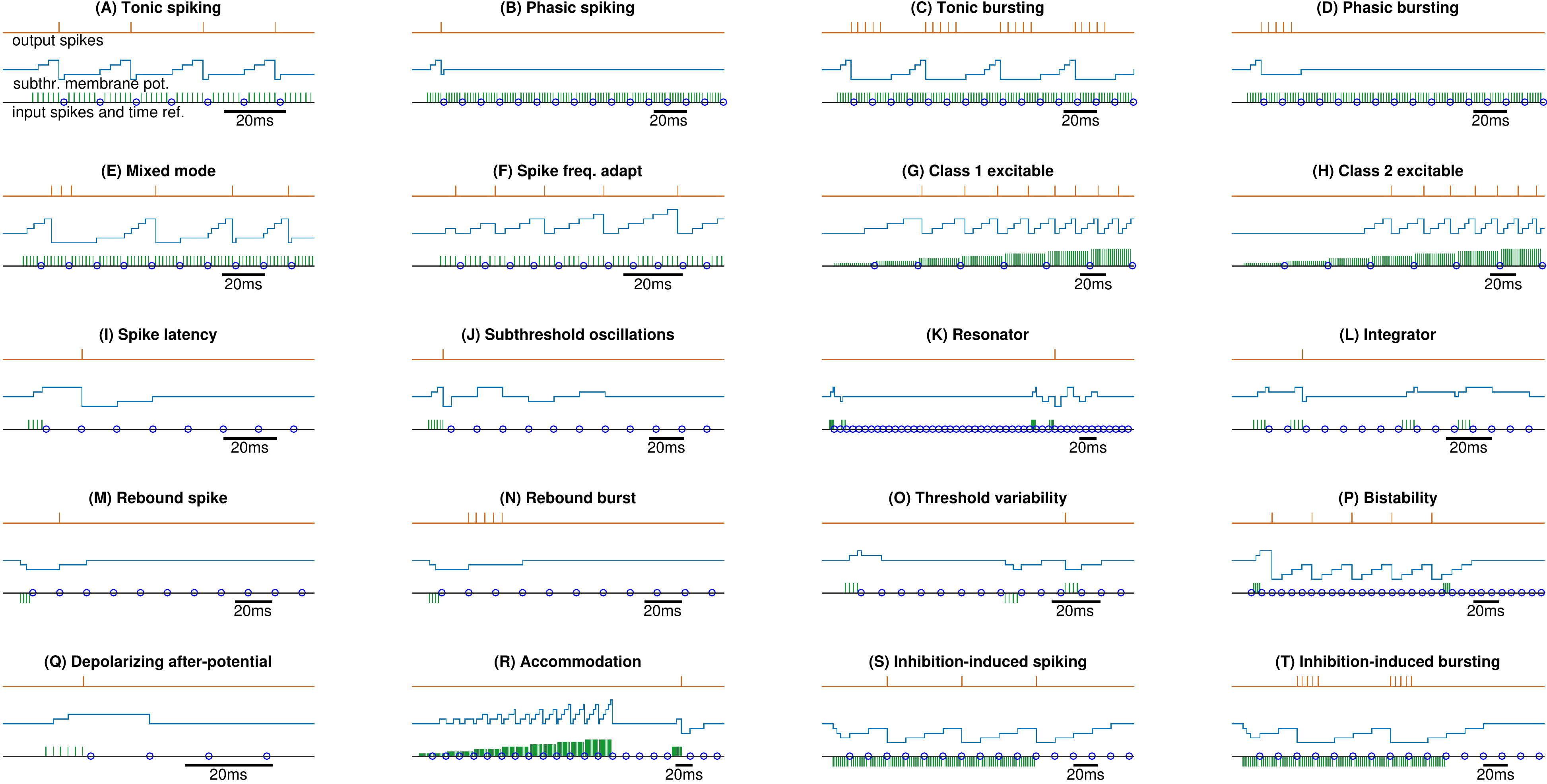}
\caption{Silicon validation of the 20 Izhikevich behaviors with biological time constant in ODIN. Scale references of 20ms are provided next to each of the 20 behaviors and closely match those of the original Izhikevich behaviors figure in \cite{Izhikevich04}.}
\label{fig_20behav}
\end{figure*}

\begin{figure*}[!t]
\centering
\noindent\includegraphics[width=0.9815\textwidth]{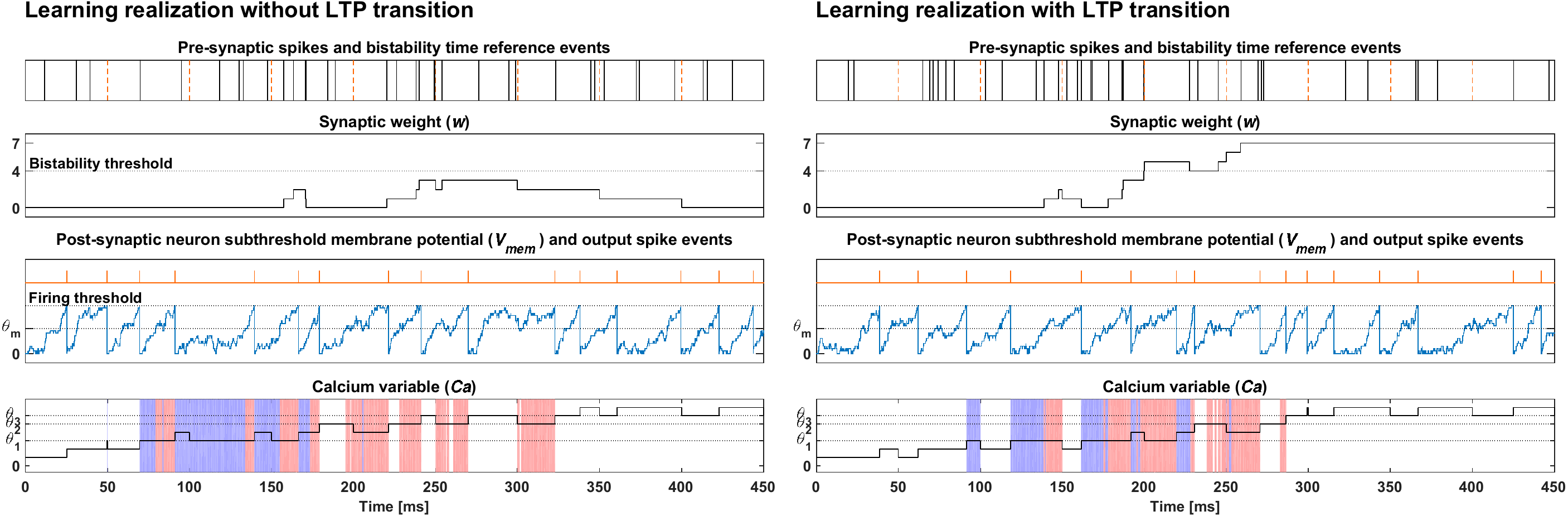}
\caption{Silicon validation of stochastic synaptic selection mechanism of the SDSP learning rule. Left: no realization of a long-term potentiation (LTP). Right: realization of an LTP after 200ms. All stimuli have been applied to a neuron in LIF configuration and generated with identical Poisson-distributed statistics: the pre-synaptic neuron fires at a rate of 70Hz, while the post-synaptic neuron fires at a rate of 40Hz. Bistability time reference events are shown in dotted orange lines at a frequency of 20Hz next to the pre-synaptic spikes. The bistability threshold at half the synaptic weight dynamic represents the value above (resp. below) which the synaptic weight increments (resp. decrements) to a high (resp. low) state upon bistability events. The SDSP conditions for weight increment and decrement from the post-synaptic neuron (\texttt{up} and \texttt{down} signals in Fig.~\ref{fig_synapseArch}) are shown next to the Calcium variable with red and blue areas, respectively. Non-colored areas of the Calcium variable are associated with stop-learning conditions.}
\label{fig_noltp_ltp}
\end{figure*}

\begin{figure*}[!t]
\centering
\vspace*{-2mm}
\noindent\includegraphics[width=0.935\textwidth]{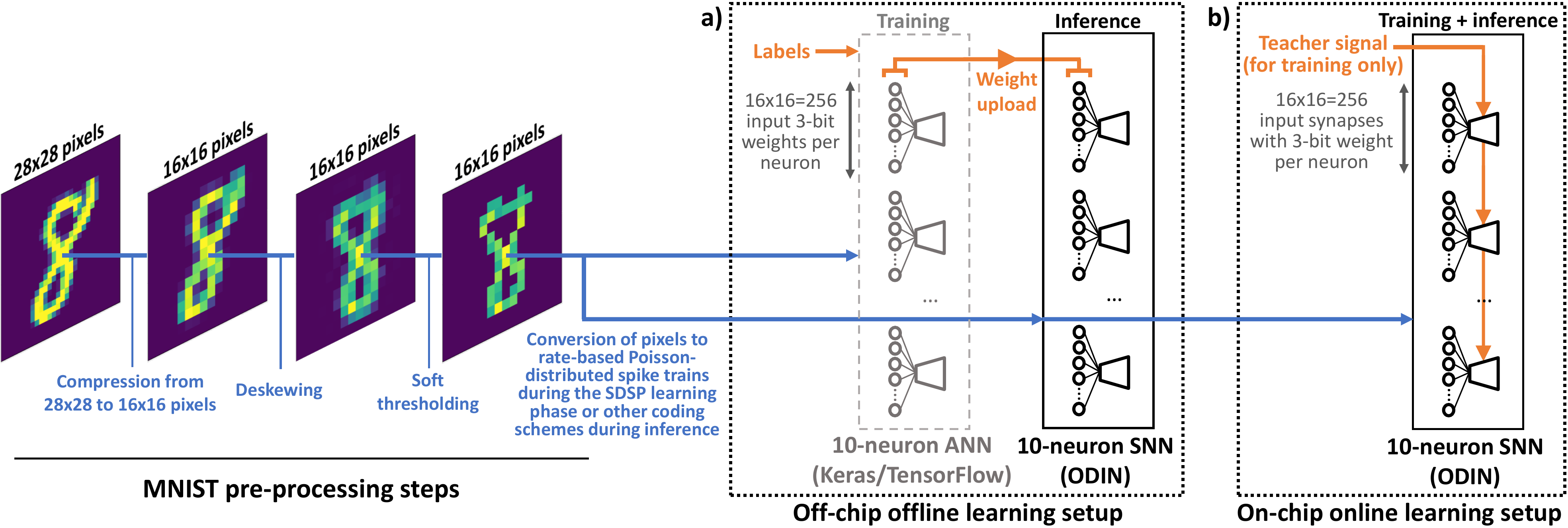}
\caption{Benchmark for testing accuracy on image classification: pre-processing steps of the MNIST dataset of handwritten digits and the two considered setups for training the weights of a LIF-based 10-neuron spiking neural network implemented in the ODIN chip. (a) Off-chip offline weight training is carried out with quantization-aware stochastic gradient descent on a 10-neuron single-layer artificial neural network (ANN) with softmax units, implemented using Keras with a TensorFlow backend. The chosen optimizer and loss function are Adam with categorical cross-entropy. (b) On-chip online teacher-based weight training with the local SDSP learning rule.}
\vspace*{-3mm}
\label{fig_MNISTflow}
\end{figure*}

\subsection{Neuron and synapse characterization} \label{ssec_neur_syn_char}

Fig.~\ref{fig_20behav} provides a silicon validation of the 20 Izhikevich behaviors in the phenomenological neuron model implemented in ODIN. Output spike patterns, subthreshold membrane potential dynamics, external input spike events and time reference events are shown at biological time constant. The time constants of the different behaviors, indicated by a 20ms scale reference next to each behavior, accurately match those of the original Izhikevich figure of the 20 behaviors in~\cite{Izhikevich04}.

A silicon validation of the SDSP online learning mechanism is shown in Fig.~\ref{fig_noltp_ltp}. The neuron is stimulated with pre-synaptic input spikes following a Poisson distribution with a 70-Hz average rate. Neuron output spikes follow a Poisson distribution with a 40-Hz average rate. Two different learning realizations follow from stimuli generated with identical statistics, illustrating the data-induced learning stochasticity and random-selection mechanism of SDSP. Indeed, on the left, no long-term potentiation (LTP) occurs, while on the right an LTP transition occurs after 200ms. In the case of supervised learning, careful sizing of teacher signals and input statistics allow controlling this probability at which a neuron will potentiate or depress the synapses in its dendritic tree~\cite{Brader07}.

\subsection{Comparison of online SDSP- and offline gradient-based supervised learning for spiking neural networks in ODIN} \label{ssec_bench}

As a mapping table bit enables or disables online learning locally in each synapse (Section~\ref{ssec_synapse}), the weights of a spiking neural network in ODIN can be learned either offline and off-chip (e.g., with gradient-descent-based algorithms) or online and on-chip with a low-cost implementation of SDSP that acts locally in the synapses. In order to identify the approach that is suitable for a given application, we compare these two learning strategies and highlight the associated tradeoffs. Given the limited hardware resources of ODIN, we chose the MNIST dataset of handwritten digits~\cite{LeCun98} as a support to discuss and quantify the comparison results on simple SNN topologies, the objective is therefore not to solve the MNIST problem with a record accuracy (already achieved with error rates as low as 0.2\%,~see~\cite{LeCun98} for a review).

The setup is presented in Fig.~\ref{fig_MNISTflow}. The pre-processing steps applied to the MNIST dataset are downsampling 28$\times$28-pixel images to a 16$\times$16 format in order to match the number of synapses per neuron available in ODIN as each pixel is mapped to an input synapse of each neuron, together with deskewing and soft thresholding, which are common operations for small networks~\cite{LeCun98}. In order to provide these images to an SNN in ODIN, one further step of converting pixel values to spikes is necessary: rate-based Poisson-distributed spike trains are used during the SDSP-based learning phase, other spike coding schemes can be explored for inference. Given that the downsampled 256-pixel MNIST images use all the 256 available neuron spike event addresses (Section~\ref{sec_arch}),  we chose to use a single-layer fully-connected network of 10 LIF neurons, one for each class of digit.

\begin{figure}[!t]
\centering
\noindent\includegraphics[width=0.96\columnwidth]{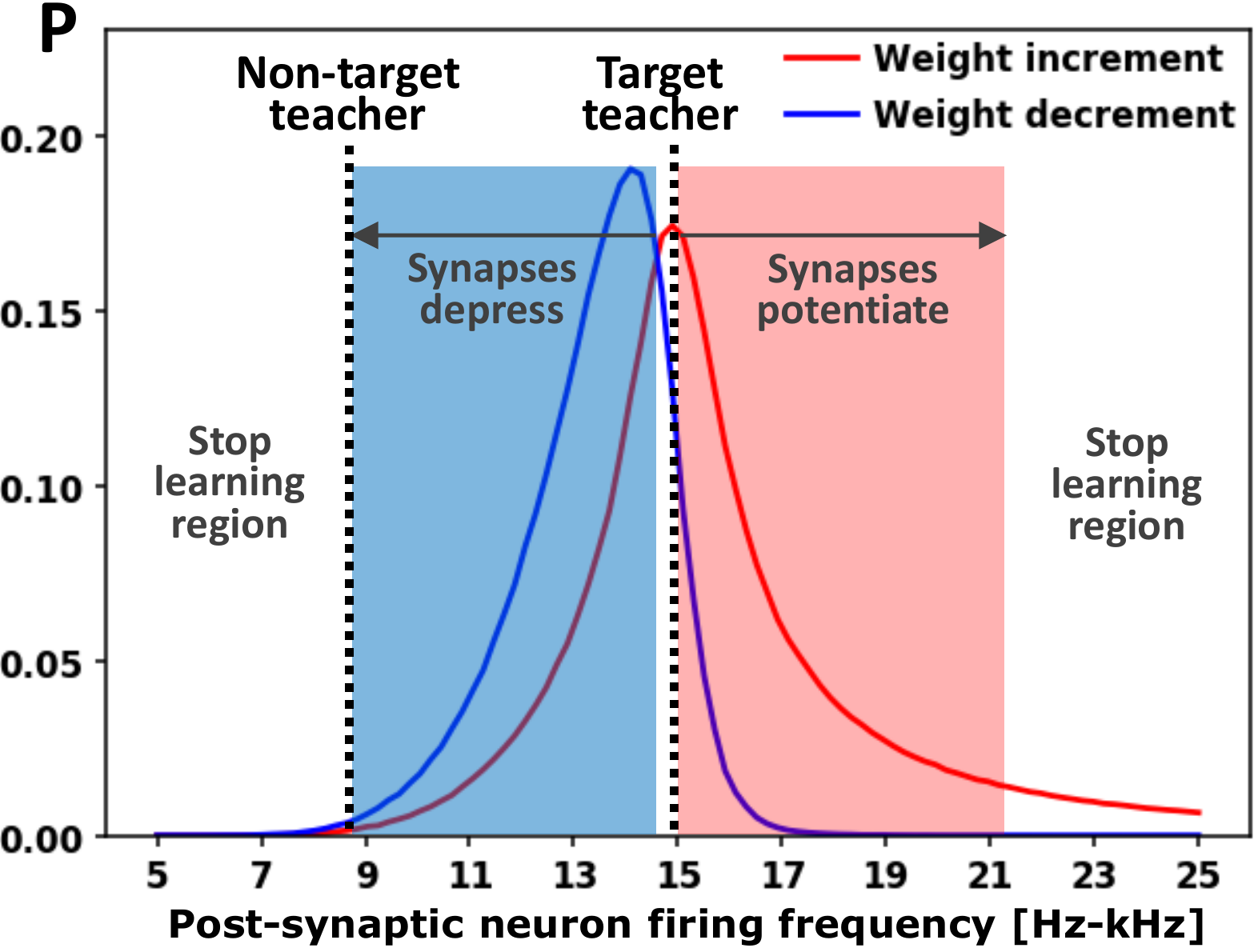}
\caption{Probability of synaptic weight increment and decrement as a function of the post-synaptic neuron firing frequency. The scale of post-synaptic neuron firing frequencies can range from biological time (Hz) to accelerated time~(kHz) if the neuron time reference events (Section~\ref{sec_arch}) are scaled accordingly. The teacher signal sets the minimum firing activity of the neuron, while red- and blue-shaded regions represent the contribution of input spike trains activity to the output neuron firing frequency.}
\label{fig_teacher}
\end{figure}

Fig.~\ref{fig_MNISTflow}a shows the setup for offline off-chip training. As the synaptic weights of ODIN have a 3-bit resolution, offline training is carried out with quantization-aware stochastic gradient descent (SGD) following~\cite{Courbariaux16}, as implemented in~\cite{Moons17} using Keras with a TensorFlow backend. The chosen optimizer is Adam, which optimizes the weights by minimizing the categorical cross-entropy loss function during several epochs, each epoch consisting in one presentation of all labeled images in the training set. The weights resulting from offline gradient-based optimization can then be uploaded directly to the LIF spiking neural network in ODIN, by simply converting the weights from a [$-4$,$3$] range as retrieved from Keras to a [$0$,$7$] range compatible with the ODIN SNN. Care should also be taken with the activation functions: softmax units were used during offline training, while the LIF neurons in the ODIN SNN behave as ReLU units in the frequency domain~\cite{Rueckauer17}. This switch from softmax to ReLU is without consequences as, during inference, the most active neuron will be the same in both cases, except if the softmax units receive a negative sum of inputs~\cite{Rueckauer17}. We verified that both the weight range conversion and the shift from softmax units to LIF neurons only incur a negligible accuracy drop compared to the theoretical accuracy predicted for the offline-trained ANN in Keras. However, while a direct weight upload is a valid approach for the simple single-layer structure used in this case, a specific weight conversion is necessary when resorting to more complex network topologies (e.g., multi-layer networks or convolutional neural networks), which can be carried out through dedicated ANN-to-SNN toolboxes, such as~\cite{Rueckauer17} or~\cite{Diehl15}.
 
Fig.~\ref{fig_MNISTflow}b shows the setup for on-chip online training with the embedded SDSP learning rule. Supervised learning with SDSP requires a teacher signal. During the training phase, the teacher signal is used to drive the neuron corresponding to the class of the currently applied digit to a high firing activity while non-target neurons are driven to low firing activities, which in turn defines the value of the neuron Calcium variable. Fig.~\ref{fig_teacher} shows how to properly set the target and non-target teacher signals with regard to the synaptic weight increment and decrement probabilities of the SDSP learning rule, which result from Eq.~(\ref{eq_sdsp}). When MNIST characters are learned (i.e. active synapses of the target neuron have potentiated), the Calcium variable crosses the stop-learning thresholds due to increased output neuron firing activity, thus preventing overfitting. During inference, no teacher signal is applied and online learning is disabled.

\begin{figure}[!t]
\centering
\noindent\includegraphics[width=0.77\columnwidth]{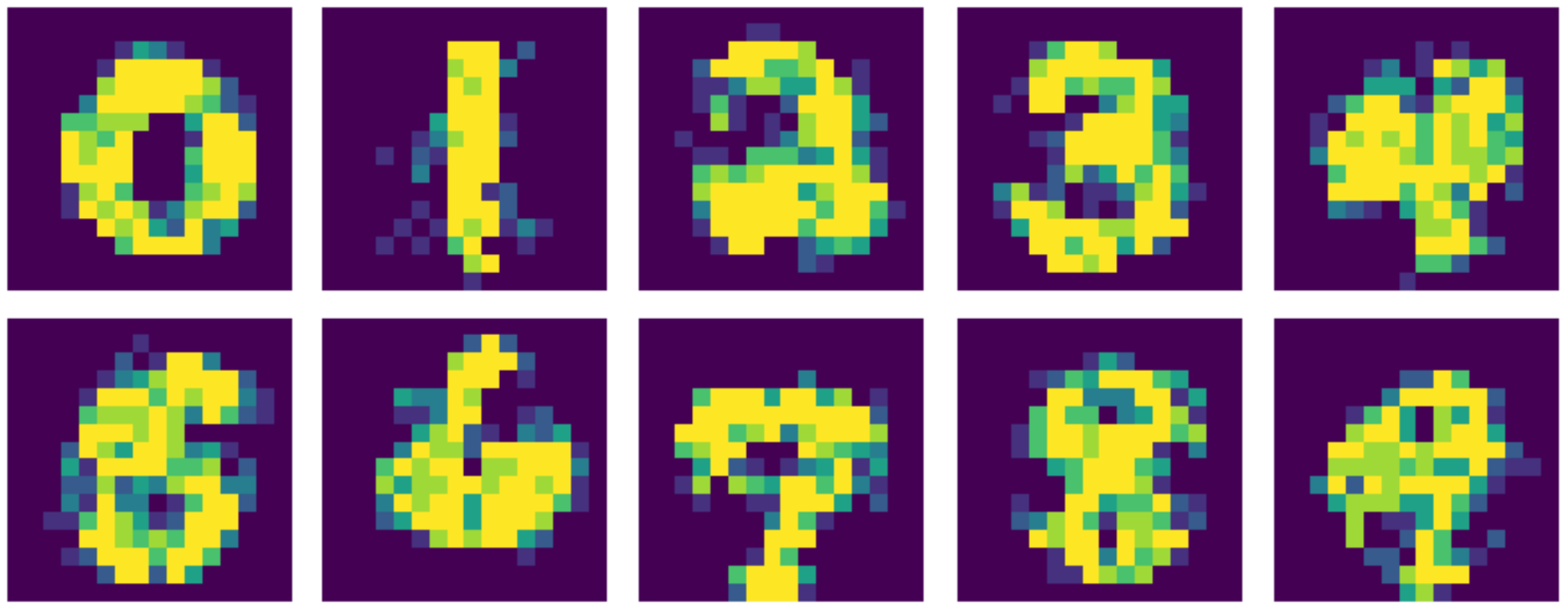}
\caption{Synaptic weights retrieved after on-chip supervised online learning of MNIST digits in ODIN with 6k training samples, leading to an accuracy of up to 85\% (rate code) or 84.5\% (rank order code) on the full 10k-sample test set. On-chip results correspond one-to-one with the simulator results.}
\label{fig_weightsSNN}
\end{figure}

\begin{table*}
\setlength\tabcolsep{4pt}
\caption{\vspace*{1mm}Chip characteristics and measured performances compared with the state of the art of spiking neuromorphic circuits.}
\label{table_SoA}
\renewcommand{\arraystretch}{1.05}
\centering
\resizebox{\textwidth}{!}{\begin{tabular}{lcccccccccccc}
\toprule%

Author & Schemmel~\cite{Schemmel10} & Benjamin~\cite{Benjamin14} & Qiao~\cite{Qiao15} & Moradi~\cite{Moradi17} & Park~\cite{Park14} & Mayr~\cite{Mayr16} & Painkras~\cite{Painkras13} & Seo~\cite{Seo11} & Kim~\cite{Kim15} & Akopyan~\cite{Akopyan15} & Davies~\cite{Davies18} & \textbf{Frenkel} \\
Publication & ISCAS, 2010 & PIEEE, 2014 & Front. NS, 2015 & TBioCAS, 2017 & BioCAS, 2014 & TBCAS, 2016 & JSSC, 2013 & CICC, 2011 & VLSI-C, 2015 & TCAD, 2015 & IEEE Micro, 2018 & \textbf{This work} \\
Chip name & HICANN & NeuroGrid & ROLLS & DYNAPs & IFAT & - & SpiNNaker & - & - & TrueNorth & Loihi & \textbf{ODIN} \\\midrule

Implementation & Mixed-signal & Mixed-signal & Mixed-signal & Mixed-signal & Mixed-signal & Mixed-signal & Digital & Digital & Digital & Digital & Digital & Digital \\

Technology & 0.18$\mu$m & 0.18$\mu$m & 0.18$\mu$m & 0.18$\mu$m & 90nm & 28nm & 0.13$\mu$m & 45nm SOI & 65nm & 28nm & 14nm FinFET & 28nm FDSOI \\\vspace*{-0.4mm}%

Neurosynaptic core area [mm$^2$] & 26.3 & 168 & 51.4 & 7.5 & 0.31 & 0.36 & 3.75 & 0.8 & 0.32 & 0.095 & 0.4 & 0.086 \\

\# Izhikevich behaviors$^\dag$ & (20) & N/A & (20) & (20) & 3 & 3 & Programmable & 3 & 3 & 11 (3 neur: 20) & (6) & 20 \\

\# neurons per core & 512 & 64k & 256 & 256 & 2k & 64 & max. 1000$^\circ$ & 256 & 64 & 256 & max. 1024 & 256 \\

Synaptic weight storage & 4-bit (SRAM) & Off-chip & Capacitor & 12-bit (CAM) & Off-chip & 4-bit (SRAM) & Off-chip & 1-bit (SRAM) & 4-,5-,14-bit (SRAM) & 1-bit (SRAM) & 1- to 9-bit (SRAM) & (3+1)-bit (SRAM)\\

Embedded online learning & STDP & No & SDSP & No & No & SDSP & Programmable & Prob. STDP & Stoch. grad. desc. & No & Programmable & SDSP \\

\# synapses per core & 112k & - & 128k & 16k & - & 8k & - & 64k & 21k & 64k & 1M to 114k (1-9 bits) & 64k \\

Time constant & Accelerated & Biological & Biological & Biological & Biological & Bio. to accel. & Bio. to accel. & Biological & N/A & Biological & N/A & Bio. to accel. \\

Neuron core density [neur/mm$^2$]$^*$ & 19.5 & 390 & 5 & 34 & 6.5k & 178 & max. 267$^\circ$ & 320 & 200 & 2.6k & max. 2.6k & 3.0k \\

Synapse core density [syn/mm$^2$]$^*$ & 4.3k & - & 2.5k & 2.1k & - & 22.2k & - & 80k & 66k & 673k & 2.5M to 285k & 741k \\

Supply voltage & 1.8V & 3.0V & 1.8V & 1.3V-1.8V & 1.2V & 0.75V, 1.0V & 1.2V & 0.53V-1.0V & 0.45V-1.0V & 0.7V-1.05V & 0.5V-1.25V & 0.55V-1.0V\\

Energy per SOP$^\ddagger$ & N/A & (941pJ)$^\blacktriangle$ & $>$77fJ$^\vartriangle$ & 134fJ$^\vartriangle$/30pJ$^\blacktriangle$ (1.3V) & 22pJ$^\blacktriangle$ & $>$850pJ$^\blacktriangle$ & $>$11.3nJ$^\vartriangle$/26.6nJ$^\blacktriangle$ & N/A & N/A & 26pJ$^\blacktriangle$ (0.775V) & $>$23.6pJ$^\vartriangle$ (0.75V) & 8.4pJ$^\vartriangle$/12.7pJ$^\blacktriangle$ (0.55V)\\

\bottomrule
\end{tabular}}

\vspace*{1.5mm}
\begin{adjustwidth}{0.3cm}{}
\begin{spacing}{0.5}
{\flushleft\fontsize{5.8pt}{6pt}\selectfont$^\dag$ By its similarity with the Izhikevich neuron model, the AdExp neuron model is believed to reach the 20 Izhikevich behaviors, but it has not been demonstrated in HICANN, ROLLS and DYNAPs. The neuron model of TrueNorth can reach 11 behaviors per neuron and 20 by combining three neurons together. The neuron model of Loihi is based on an LIF model to which threshold adaptation is added: the neuron should therefore reach up to 6 Izhikevich behaviors, although it has not been demonstrated.}
\end{spacing}
\vspace*{2.7mm}
\end{adjustwidth}
\begin{adjustwidth}{0.3cm}{}
\begin{spacing}{0.5}
{\flushleft\fontsize{5.8pt}{6pt}\selectfont$^\circ$ Experiment 1 reported in Table~III from~\cite{Painkras13}, is considered as a best-case neuron density: 1000 simple LIF neuron models are implemented per core, each firing at a low frequency.}
\end{spacing}
\vspace*{2.6mm}
\end{adjustwidth}
\begin{adjustwidth}{0.3cm}{}
\begin{spacing}{0.5}
{\flushleft\fontsize{5.8pt}{6pt}\selectfont$^*$ Neuron (resp. synapse) core densities are computed by dividing the number of neurons (resp. synapses) per neurosynaptic core by the neurosynaptic core area. Regarding the synapse core density, NeuroGrid, IFAT and SpiNNaker use an off-chip memory to store synaptic data. As the synapse core density cannot be extracted when off-chip resources are involved, no synapse core density values are reported for these chips.}
\end{spacing}
\end{adjustwidth}
\vspace*{-1.7mm}
\begin{adjustwidth}{0.3cm}{}
\begin{spacing}{0.5}
{\flushleft\fontsize{5.8pt}{6pt}\selectfont$^\ddagger$ The synaptic operation energy measurements reported for the different chips do not follow a standardized measurement process and are provided only for reference. There are two main categories for energy measurements in neuromorphic chips. On the one hand, incremental values (denoted with $^\vartriangle$) describe the amount of energy paid per each additional SOP computation, they are measured by subtracting the leakage and idle power consumption of the chip, as in Eq.~(\ref{eq_power}), although the exact power contributions taken into account in the SOP energy vary across chips. On the other hand, global values (denoted with $^\blacktriangle$) are obtained by dividing the total chip power consumption by the SOP rate, as in Eq.~(\ref{eq_powerTot}). The conditions under which all of these measurements have been done can be found hereafter. For NeuroGrid, a SOP energy of 941pJ is reported for a network of 16 Neurocore chips (1M neurons, 8B synapses, 413k spikes/s): it is a board-level measurement, no chip-level measurement is provided~\cite{Benjamin14}. For ROLLS, the measured SOP energy of 77fJ is reported in~\cite{Indiveri15b}, it accounts for a point-to-point synaptic input event and includes the contribution of weight adaptation and digital-to-analog conversion, it represents a lower bound as it does not account for synaptic event broadcasting. For DYNAPs, the measured SOP energy of 134fJ at 1.3V is also reported in~\cite{Indiveri15b} while the global SOP energy of 30pJ can be estimated from~\cite{Moradi17} using the measured 800-$\mu$W power consumption with all 1k neurons spiking at 100Hz with 25\% connectivity (26.2MSOP/s), excluding the synaptic input currents. For IFAT, the SOP energy of 22pJ is extracted by measuring the chip power consumption when operated at the peak rate of 73M synaptic events/s~\cite{Park14}. In the chip of Mayr~\textit{et~al.}, the SOP energy of 850pJ represents a lower bound extracted from the chip power consumption, estimated by considering the synaptic weights at half their dynamic at maximum operating frequency~\cite{Mayr16}. For SpiNNaker, an incremental SOP energy of 11.3nJ is measured in~\cite{Stromatias15}, a global SOP energy of 26.6nJ at the maximum SOP rate of 16.56MSOP/s can be estimated by taking into account the leakage and idle power; both values represent a lower bound as the energy cost of neuron updates is not included. In the chip of Kim~\textit{et~al.}, an energy per pixel of 5.7pJ at 0.45V is provided from a chip-level power measurement during inference (i.e. excluding the learning co-processor), but no information is provided on how to relate this number to the energy per SOP. For TrueNorth, the measured SOP energy of 26pJ at 0.775V is reported in~\cite{Merolla14}, it is extracted by measuring the chip power consumption when all neurons fire at 20Hz with 128 active synapses. For Loihi, a minimum SOP energy of 23.6pJ at 0.75V is extracted from pre-silicon SDF and SPICE simulations, in accordance with early post-silicon characterization~\cite{Davies18}; it represents a lower bound as it includes only the contribution of the synaptic operation, without taking into account the cost of neuron update and learning engine update. For ODIN, the detailed measurement process is described in Section~\ref{ssec_chip_spef}.}
\end{spacing}
\end{adjustwidth}
\end{table*}

The maximum attainable accuracy of the 10-neuron 3-bit-weight SNN trained online with SDSP saturates at 85\% for training set sizes equal or higher than 6k samples, which corresponds to the synaptic weights retrieved from ODIN after on-chip online learning shown in Fig.~\ref{fig_weightsSNN}. The 6k training samples were presented only once to the network, increasing the training set size or the number of presentations does not further improve the performance of the final classifier. At 0.55V, ODIN consumes 105nJ/sample during the learning phase relying on the SDSP-based supervised learning setup described previously, which includes contributions from both the teacher signal (95nJ) and the input character (10nJ). A higher accuracy of 91.9\% can be reached in ODIN using offline SGD-based training using all the 60k MNIST training samples with 100 epochs (the theoretical accuracy obtained in Keras, before the weight upload to ODIN, was 92.5\%).

\begin{figure}[!t]
\centering
\vspace*{1mm}
\noindent\includegraphics[width=1.0\columnwidth]{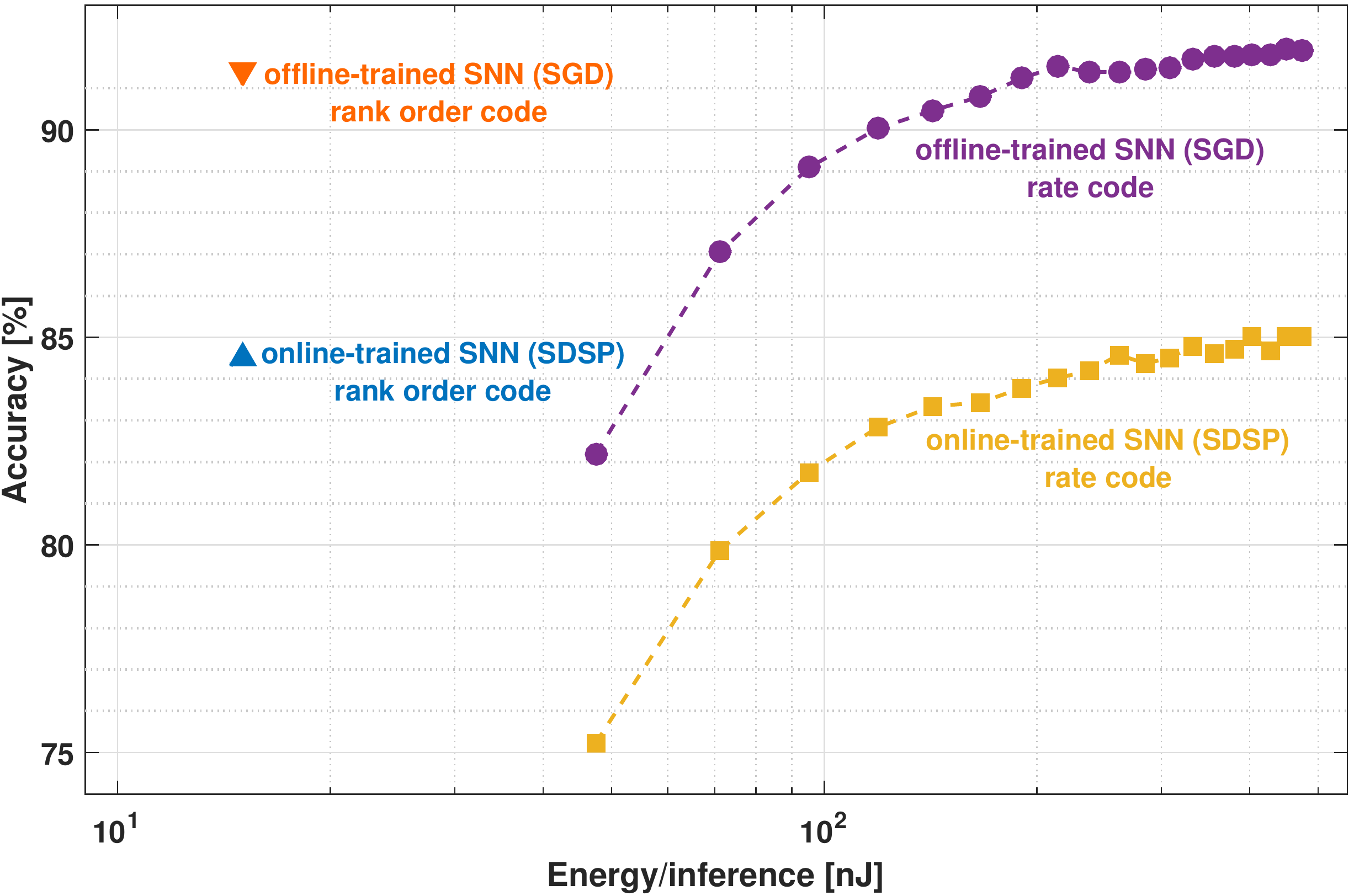}
\caption{Accuracy vs. energy per inference tradeoff for the rank order code and the rate code on the MNIST dataset. For the rate code, this tradeoff depends on the character presentation time: the longer the presentation, the higher the accumulated statistics and the consumed energy. For the rank order code, the character presentation is stopped as soon as the decision is made by a single output spike, leading to an efficient spike use. Results are provided for the ODIN chip using either weights learned off-chip offline with stochastic gradient descent (SGD) or weights learned on-chip online with SDSP, following the setup from Fig.~\ref{fig_MNISTflow}.}
\label{fig_rankVSrate}
\end{figure}

During the inference phase, different coding schemes can be explored for the weights learned either online with SDSP or offline with SGD, resulting in different tradeoffs between accuracy and energy per inference, as shown in Fig.~\ref{fig_rankVSrate}. On the one hand, the rate code is a standard spike coding approach in which pixel values are encoded into spike rates~\cite{Thorpe01}, the output neuron that spikes the most defines the inferred class. In this case, ODIN consumes 404nJ/inference after on-chip learning with 85\% accuracy and 451 nJ/inference after off-chip learning with 91.9\% accuracy. Input characters must be presented for a much longer time than during the on-chip learning phase in order to discriminate characters, sufficient statistics need to be accumulated in the output neurons. The rate code is thus inefficient in its spike use. On the other hand, the rank order code gives a promising approach for a low-power spike coding alternative that is easy to use with LIF neurons~\cite{Thorpe01}. In the rank order code, each character is provided to the ODIN SNN as a sequence of successive spikes where no relative timing is involved, each pixel spikes once and all pixels are arranged in order of decreasing value in the sequence. The sequence is then presented repeatedly to the ODIN SNN until one output neuron spikes, this first-spiking neuron defines the inferred class. In this case, ODIN consumes only 15nJ/inference for 84.5\% accuracy after on-chip learning (resp. 91.4\% accuracy after off-chip learning), which demonstrates a high efficiency (e.g.,~see~\cite{Whatmough17} for a review of energy per classification of previously-proposed ANNs, CNNs and SNNs on the original 28$\times$28 MNIST dataset). To the best of our knowledge, this is the first time MNIST classification is demonstrated with on-chip SDSP-based learning. Therefore, Fig.~\ref{fig_rankVSrate} highlights that, on the one hand, the SDSP learning rule allows to carry out spiking neural network training on-chip and on-line with low hardware resources at the cost of a limited accuracy degradation. It corresponds to applications that are constrained in power and resources both during the training phase and the inference phase and where data is presented on-the-fly to the classifier (Section~\ref{sec_intro}). On the other hand, resorting to the use of an off-chip learning engine is the optimum approach for applications that are not power- or resource-constrained during the training phase: it allows reaching a higher accuracy while keeping power efficiency during the inference phase.

Finally, given more neuron resources (e.g., by connecting several ODIN chips together), a higher accuracy on MNIST could be attained with more complex network topologies. In the case of online SDSP-based learning, two-layer reservoirs and unsupervised WTA-based networks are interesting topologies. Both of these network topologies have been recently studied for subthreshold analog implementations of SDSP (e.g., see~\cite{Corradi15} and~\cite{Kreiser17}, respectively), we intend to explore them in future work. If high accuracy is needed through global gradient-based optimization with deep multi-layer networks, the weights from an offline-learned ANN topology can be mapped to an SNN in ODIN using dedicated ANN-to-SNN toolboxes (e.g., \cite{Rueckauer17} or \cite{Diehl15}). As the synaptic weights of ODIN can be individually configured to be static or plastic, hybrid networks with SDSP-based plastic and pre-trained static layers can also be investigated, so that a basic learning architecture can be adapted online to the environment.

\section{Discussion} \label{sec_disc}

A performance and specification summary of state-of-the-art neuromorphic chips is provided in Table~\ref{table_SoA}. Mixed-signal designs with core analog neuron and synapse computation and high-speed digital periphery are grouped on the left~\cite{Schemmel10,Benjamin14,Qiao15,Moradi17,Park14,Mayr16}, digital designs are grouped together with ODIN on the right~\cite{Painkras13,Seo11,Kim15,Akopyan15,Davies18}. Toward efficient spiking neuromorphic experimentation platforms, the key figures of merit are neuron versatility (i.e.~Izhikevich behavior repertoire), synaptic plasticity,~\mbox{density and energy per SOP.}

Several chips in Table~\ref{table_SoA} embed specific routing schemes to allow for large-scale integration of interconnected chips, it is the case for HICANN, NeuroGrid, DYNAPs, SpiNNaker, TrueNorth and Loihi, while a hierarchical routing AER topology has been proposed recently for IFAT in~\cite{Park17}. In order to carry out fair comparisons, when these chips are composed of several neurosynaptic cores, we reported in Table~\ref{table_SoA} the density data associated to a single neurosynaptic core. The other chips compared in Table~\ref{table_SoA}, including ODIN, consist of a single neurosynaptic core. Large-scale interconnection can be achieved using the standard input and output AER interfaces, but this connection scheme requires an external routing table in order to define the inter-chip connectivity. The scope of ODIN lies in the design of a power- and area-efficient neurosynaptic core, we expect in future work to explore hierarchical event routing infrastructures in order to move the ODIN~\mbox{neurosynaptic core to efficient large-scale integration.} 

Among all SNNs, the 28nm IBM TrueNorth chip~\cite{Akopyan15} previously had the best neuron and synapse densities, well beyond those of all mixed-signal approaches proposed to date. Compared to ODIN, its neurosynaptic cores have identical numbers of neurons and synapses and both chips are in 28nm CMOS technology, which allows direct comparison. While TrueNorth does not embed synaptic plasticity, we show with ODIN that it is possible to quadruple the number of bits per synapse and to add online learning while slightly reducing the overall area,~\mbox{thus improving overall neuron and synapse densities}.

The 14nm Intel Loihi chip has recently been proposed~\cite{Davies18} and embeds a configurable spike-timing-based learning rule. Loihi is also an experimentation platform and has specific features. While the neuron Calcium variable in ODIN corresponds to an SDSP eligibility trace with configurable time constant through Calcium leakage, Loihi offers several types of eligibility traces (e.g., multiple spike traces and reward traces) with extended configurability in the context of its programmable learning rule. Its axonal and refractory delays, stochasticity and threshold adaptation for homeostasis can be captured in ODIN by the stochastic transitions of SDSP~(Fig.~\ref{fig_noltp_ltp}) and by refractory, spike latency and threshold variability behaviors of the phenomenological Izhikevich neuron. Loihi also allows for dendritic tree computation with multi-compartment neurons, while ODIN further extends the behavior repertoire. Loihi has configurable synaptic fan-in/resolution tradeoff, but despite the fact that Loihi is implemented in a more advanced 14-nm FinFET technology node, Table~\ref{table_SoA} shows that ODIN compares favorably to Loihi in neuron core density (i.e. 3k neurons/mm$^2$ for ODIN, max. 2.6k for Loihi) and synapse core density (i.e. 741k synapses/mm$^2$ for ODIN, 625k for Loihi with 4-bit synapses).

Regarding power consumption, as indicated in Table~\ref{table_SoA}, the SOP energies reported in the state of the art are provided for reference and should not be compared directly due to non-standardized measurement processes across the different chips. However, it allows to extract orders of magnitude. The flexibility-power tradeoff of the SpiNNaker software approach appears clearly: high programmability for both neuron and synapse models leads to a global energy per SOP of 26.6nJ. The next SpiNNaker generation might improve this tradeoff through advanced power reduction techniques and dedicated hardware accelerators~(e.g.,~\cite{Hoppner17,Partzsch17}). For ODIN, supply voltage scaling down to 0.55V in 28nm FDSOI CMOS results in a minimum global energy per SOP of 12.7pJ at the maximum SOP rate. Table~\ref{table_SoA} also shows that the hundred-fJ incremental energy of subthreshold analog designs make them appear as particularly energy-efficient given that they exclude contributions of leakage, idle power and network operation. However, when taking these elements into account, the high ratio between global and incremental energies per SOP of DYNAPs shows that these designs do not scale efficiently.

\section{Conclusion} \label{sec_ccl}

In this work, we demonstrated ODIN, a digital spiking neuromorphic processor implemented in 28nm FDSOI CMOS technology. It embeds 256 neurons and 64k synapses in an area of only 0.086mm$^2$ and emulates each of the 20 Izhikevich behaviors. The SDSP learning rule with overfitting-prevention mechanism is embedded in all synapses at high density with 0.68$\mu$m$^2$ per 4-bit synapse with embedded online learning. ODIN has the highest neuron and synapse densities among all mixed-signal and digital SNNs proposed to date, while exhibiting a global energy per SOP down to 12.7pJ. Using a single-layer spiking neural network on ODIN, we show that SDSP-based on-chip online learning allows training approximate classifiers on the MNIST classification task, which is tailored to applications that are constrained in power and resources during the training phase. Offline learning allows reaching a higher accuracy if the target application does not have stringent power or resource constraints during the learning phase. Both the online and offline training approaches leverage the energy efficiency of ODIN with only 15nJ per classification during the inference phase, with accuracies of 84.5\% and 91.4\%~respectively.

These results demonstrate that a deeply-scaled digital approach can be leveraged for high-density and low-power spiking neural network implementation.

\section*{Acknowledgment}

The authors would like to thank Giacomo Indiveri and Michel Verleysen for fruitful discussions, Thomas Haine for the design of the clock generator, the participants of the Capo Caccia Neuromorphic Engineering workshop for valuable feedback and the anonymous reviewers for their help in improving the contents and clarity of the manuscript.

C.~Frenkel is with Universit\'e catholique de Louvain as a Research Fellow from the National Foundation for Scientific Research (FNRS) of Belgium.

\vskip 34pt plus -1fil

\begin{IEEEbiography}
	[{\includegraphics[width=1in,height=1.25in,clip,keepaspectratio]{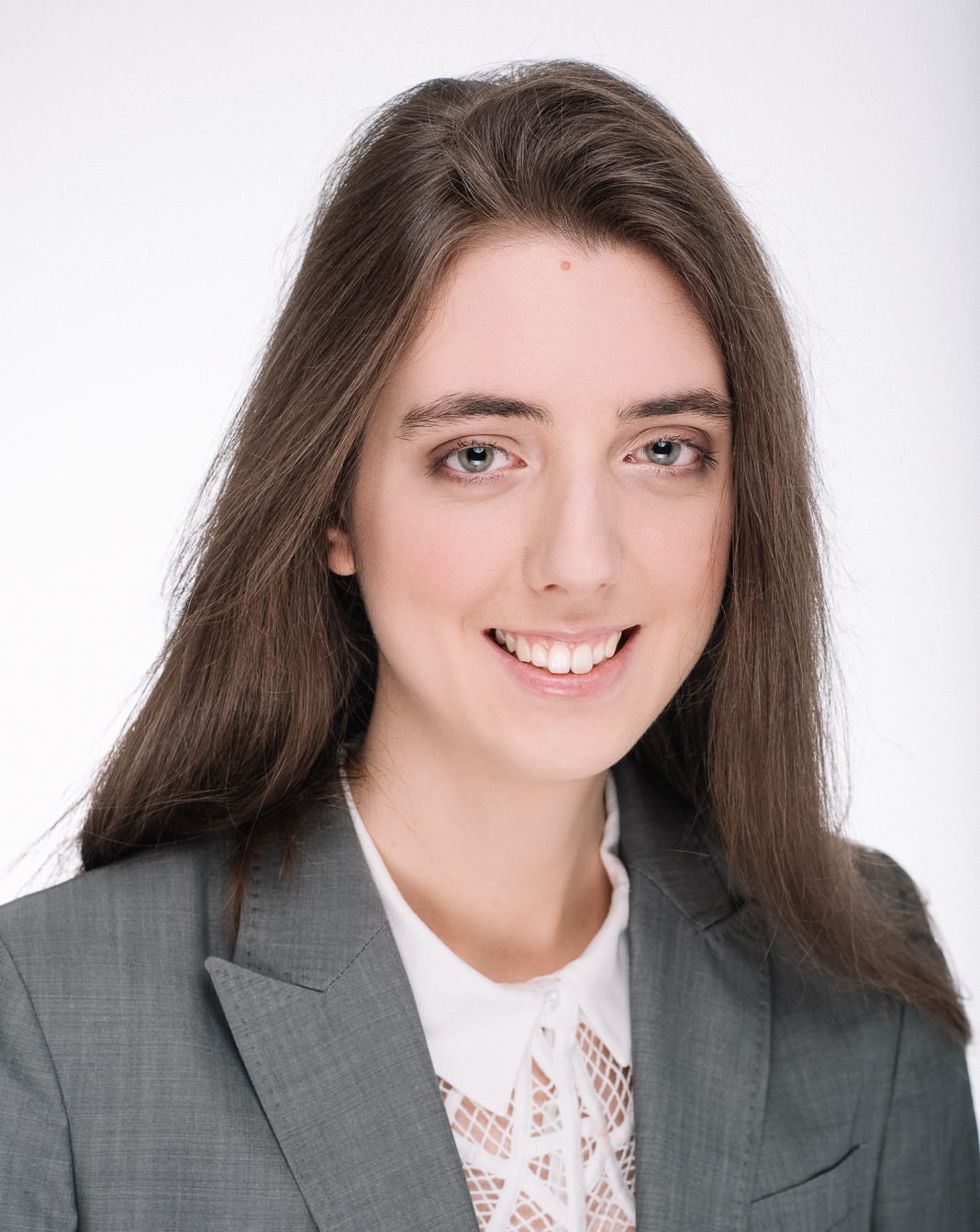}}]{Charlotte Frenkel}
(S'15) received the M.Sc. degree (\textit{summa cum laude}) in Electromechanical Engineering from Universit\'e catholique de Louvain (UCL), Louvain-la-Neuve, Belgium, in 2015. She is currently working toward the Ph.D. degree as a Research Fellow of the National Foundation for Scientific Research (FNRS) of Belgium, under the supervision of Prof. D. Bol and Prof. J.-D. Legat.

Her current research focuses on the design of low-power and high-density neuromorphic circuits as efficient non-von Neumann architectures for real-time recognition and online learning.

Ms. Frenkel serves as a reviewer for the IEEE Transactions on Neural Networks and Learning Systems journal and for IEEE S3S, ISCAS and BioCAS conferences.
\end{IEEEbiography}

\vskip 36.5pt plus -1fil

\begin{IEEEbiography}
	[{\includegraphics[width=1in,height=1.25in,clip,keepaspectratio]{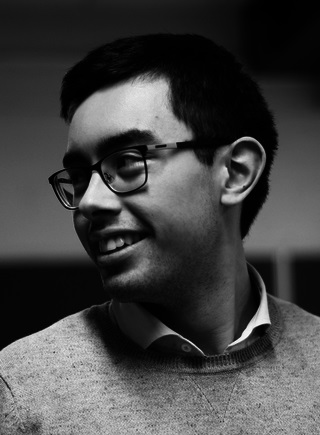}}]{Martin Lefebvre}
(S'17) received the M.Sc. degree (\textit{summa cum laude}) in electromechanical engineering from the Universit\'e catholique de Louvain (UCL), Louvain-la-Neuve, Belgium, in 2017, where he is currently pursuing the Ph.D. degree, under the supervision of Prof. D. Bol and Prof. L. Jacques.

His current research interests include the design of low-power mixed-signal circuits as efficient hardware implementations for machine learning and image processing algorithms.
\end{IEEEbiography}

\newpage

\begin{IEEEbiography}
	[{\includegraphics[width=1in,height=1.25in,clip,keepaspectratio]{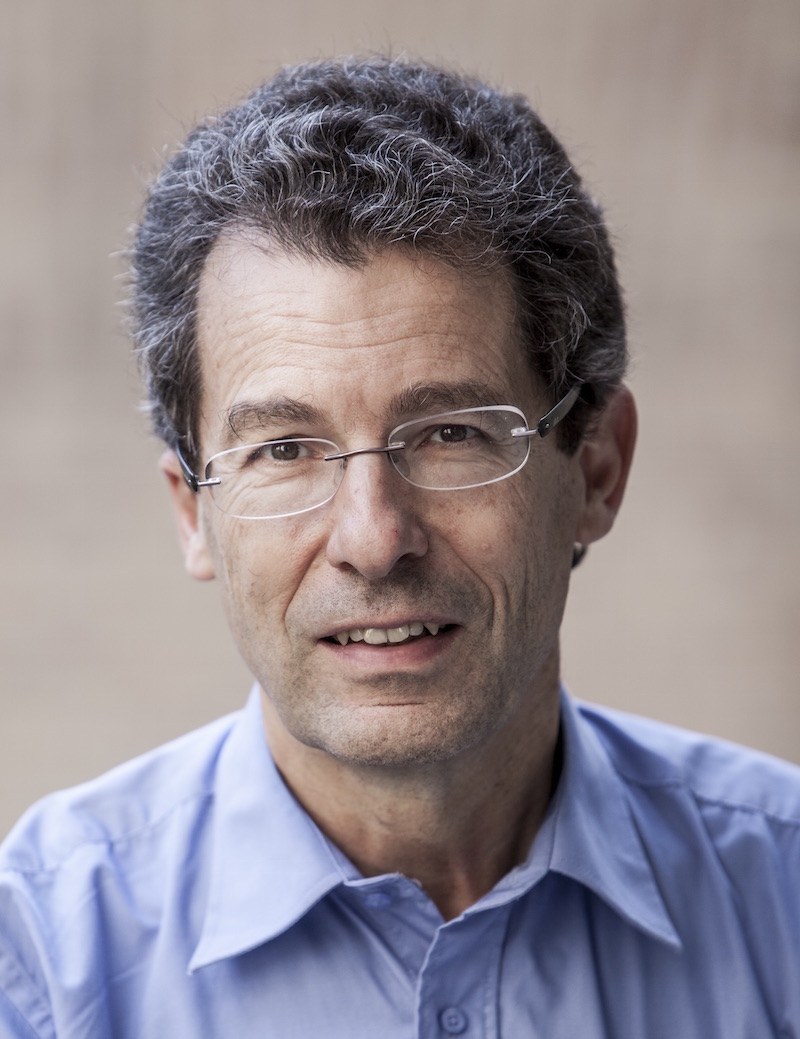}}]{Jean-Didier Legat}
(S'79-M'85-SM'17) received his engineering and PhD degrees in microelectronics from the Universit\'e catholique de Louvain, Louvain-la-Neuve, Belgium in 1981 and 1987, respectively.

From 1987 to 1990, he was with Image Recognition Integrated Systems (I.R.I.S.), a new company specialised in optical character recognition and automatic document processing. Jean-Didier Legat was co-founder and Vice-President of I.R.I.S. In October 1990, he came back to the UCL Microelectronics Laboratory. He is presently full Professor. From 2003 to 2008, he was the Dean of the Louvain School of Engineering. Currently, he is Senior Advisor to the President for Technology Transfer and Head of the ICTEAM Research Institute. His current interests are processor architecture, low-power digital integrated circuit, real-time embedded systems, mixed-signal design and hardware-software codesign for reconfigurable systems. He has been an author or co-author of more than 200 publications in the field of microelectronics, low-power digital circuits, computer architecture, digital signal processing, computer vision and pattern recognition.
\end{IEEEbiography}

\vskip 0pt plus -1fil

\begin{IEEEbiography}
	[{\includegraphics[width=1in,height=1.25in,clip,keepaspectratio]{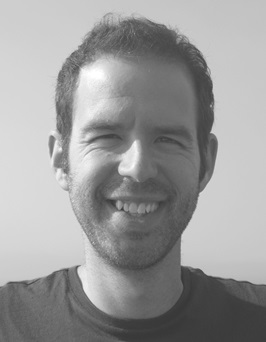}}]{David Bol}
(S'07-M'09-SM'18) received the M.Sc. degree in Electromechanical Engineering and the Ph.D degree in Engineering Science from Universit\'e catholique de Louvain (UCL), Louvain-la-Neuve, Belgium in 2004 and 2008, respectively. In 2005, he was a visiting Ph.D student at the CNM National Centre for Microelectronics, Sevilla, Spain, in advanced logic design. In 2009, he was a postdoctoral researcher at intoPIX, Louvain-la-Neuve, Belgium, in low-power design for JPEG2000 image processing. In 2010, he was a visiting postdoctoral researcher at the UC Berkeley Laboratory for Manufacturing and Sustainability, Berkeley, CA, in life-cycle assessment of the semiconductor environmental impact. He is now an assistant professor at UCL. In 2015, he participated to the creation of e-peas semiconductors, Louvain-la-Neuve, Belgium.

Prof. Bol leads with Prof. Denis Flandre the Electronic Circuits and Systems (ECS) research group focused on ultra-low-power design of integrated circuits for the IoT including computing, power management, sensing and RF communications with focuses on technology/circuit interaction in nanometer CMOS nodes, variability mitigation, mixed-signal SoC architecture and implementation. He gives four M.Sc. courses in Electrical Engineering at UCL on digital, analog and mixed-signal integrated circuits and systems as well as sensors.

Prof. Bol has authored or co-authored more than 100 technical papers and conference contributions and holds three delivered patents. He (co-)received three Best Paper/Poster/Design Awards in IEEE conferences (ICCD 2008, SOI Conf. 2008, FTFC 2014). He also serves as an editor for MDPI J. Low-Power Electronics and Applications, as a TPC member of IEEE SubVt/S3S conference and as a reviewer for various journals and conferences such as IEEE J. of Solid-State Circuits, IEEE Trans. on VLSI Syst., IEEE Trans. on Circuits and Syst. I/II. Since 2008, he presented several invited papers and keynote tutorials in international conferences including a forum presentation at IEEE ISSCC 2018.
\end{IEEEbiography}


\begin{thebibliography}{1}


\bibitem{Bol15}
D. Bol, G. de Streel and D. Flandre, ``Can we connect trillions of IoT sensors in a sustainable way? A technology/circuit perspective,'' \textit{Proc. of IEEE SOI-3D-Subthreshold Microelectronics Technology Unified Conference (S3S)}, 2015.

\bibitem{Horowitz14}
M. Horowitz, ``1.1 Computing's energy problem (and what we can do about it),'' \textit{Proc. of IEEE International Solid-State Circuits Conference Digest of Technical Papers (ISSCC}), pp. 10-14, 2014.

\bibitem{Sandin14}
F. Sandin et al., ``Concept learning in neuromorphic vision systems: What can we learn from insects?,'' \textit{Journal of Software Engineering and Applications}, vol. 7, no. 5, pp. 387-395, 2014.

\bibitem{Chittka09}
L. Chittka and J. Niven, ``Are bigger brains better?,'' \textit{Current Biology}, vol. 19, no. 21, pp. R995-R1008, 2009.

\bibitem{Liu10}
S.-C. Liu and T. Delbruck, ``Neuromorphic sensory systems,'' \textit{Current Opinion in Neurobiology}, vol. 20, no. 3, pp. 288-295, 2010.

\bibitem{Indiveri15a}
G. Indiveri and S.-C. Liu, ``Memory and information processing in neuromorphic systems,'' \textit{Proceedings of the IEEE}, vol. 103, no. 8, pp.~1379-1397, 2015.

\bibitem{Sandamirskaya14}
Y. Sandamirskaya, ``Dynamic neural fields as a step toward cognitive neuromorphic architectures,'' \textit{Frontiers in Neuroscience}, vol. 7, p. 276, 2014.

\bibitem{Milde17}
M. B. Milde et al., ``Obstacle avoidance and target acquisition for robot navigation using a mixed signal analog/digital neuromorphic processing system,'' \textit{Frontiers in Neurorobotics}, vol. 11, p. 28, 2017.

\bibitem{Corradi15}
F. Corradi and G. Indiveri, ``A neuromorphic event-based neural recording system for smart brain-machine-interfaces,'' \textit{IEEE Transactions on Biomedical Circuits and Systems}, vol. 9, no. 5, pp. 699-709, 2015.

\bibitem{Boi16}
F. Boi et al., ``A bidirectional brain-machine interface featuring a neuromorphic hardware decoder,'' \textit{Frontiers in Neuroscience}, vol. 10, p.~563, 2016.

\bibitem{Vogelstein08}
R. J. Vogelstein et al., ``A silicon central pattern generator controls locomotion in vivo,'' \textit{IEEE Transactions on Biomedical Circuits and Systems}, vol. 2, no. 3, pp. 212-222, 2008.

\bibitem{George15}
R. George et al., ``Event-based softcore processor in a biohybrid setup applied to structural plasticity,'' \textit{Proc. of IEEE International Conference on Event-based Control, Communication, and Signal Processing (EBCCSP)}, 2015.

\bibitem{Indiveri09}
G. Indiveri, E. Chicca and R. J. Douglas, ``Artificial cognitive systems: from VLSI networks of spiking neurons to neuromorphic cognition,'' \textit{Cognitive Computation}, vol. 1, no. 2, pp. 119-127, 2009.

\bibitem{Izhikevich04}
E. M. Izhikevich, ``Which model to use for cortical spiking neurons?,'' \textit{IEEE Transactions on Neural Networks}, vol. 15, no. 5, pp. 1063-1070, 2004.

\bibitem{Azghadi14}
M. R. Azghadi et al., ``Spike-based synaptic plasticity in silicon: design, implementation, application, and challenges,'' \textit{Proceedings of the IEEE}, vol. 102, no. 5, pp. 717-737, 2014.

\bibitem{Painkras13}
E. Painkras et al., ``SpiNNaker: A 1-W 18-core system-on-chip for massively-parallel neural network simulation,'' \textit{IEEE Journal of Solid-State Circuits}, vol. 48, no. 8, pp. 1943-1953, 2013.

\bibitem{Cassidy13}
A. Cassidy, J. Georgiou and A. G. Andreou, ``Design of silicon brains in the nano-CMOS era: Spiking neurons, learning synapses and neural architecture optimization,'' \textit{Neural Networks}, vol.~45, pp.~4-26, 2013.

\bibitem{Neil14}
D. Neil and S.-C. Liu, ``Minitaur, an event-driven FPGA-based spiking network accelerator,'' \textit{IEEE Transactions on Very Large Scale Integration (VLSI) Systems}, vol.~22, no.~12, pp.~2621-2628, 2014.

\bibitem{Luo16}
J. Luo et al., ``Real-Time Simulation of Passage-of-Time Encoding in Cerebellum Using a Scalable FPGA-Based System,'' \textit{IEEE Transactions on Biomedical Circuits and Systems}, vol.~10, no.~3, pp.~742-753, 2016.

\bibitem{Yang18}
S. Yang et al., ``Real-Time Neuromorphic System for Large-Scale Conductance-Based Spiking Neural Networks,'' \textit{IEEE Transactions on Cybernetics}, 2018.

\bibitem{Poon11}
C. S. Poon and K. Zhou, ``Neuromorphic silicon neurons and large-scale neural networks: challenges and opportunities,'' \textit{Frontiers in Neuroscience}, vol. 5, p. 108, 2011.

\bibitem{Mead89}
C. Mead, \textit{Analog VLSI and Neural Systems.} Reading, MA: Addison-Wesley, 1989.

\bibitem{Qiao15}
N. Qiao et al., ``A reconfigurable on-line learning spiking neuromorphic processor comprising 256 neurons and 128K synapses,'' \textit{Frontiers in Neuroscience}, vol. 9, no. 141, 2015.%

\bibitem{Moradi17}
S. Moradi et al., ``A scalable multicore architecture with heterogeneous memory structures for Dynamic Neuromorphic Asynchronous Processors (DYNAPs),'' \textit{IEEE Transactions on Biomedical Circuits and Systems}, vol.~12, no. 1, pp. 106-122, 2017.

\bibitem{Schemmel10}
J. Schemmel et al., ``A wafer-scale neuromorphic hardware system for large-scale neural modeling,'' \textit{Proc. of IEEE International Symposium on Circuits and Systems (ISCAS)}, pp. 1947-1950, 2010.%

\bibitem{Noack15}
M. Noack et al., ``Switched-capacitor realization of presynaptic short-term-plasticity and stop-learning synapses in 28 nm CMOS,'' \textit{Frontiers in neuroscience}, vol.~9, p.~10, 2015.

\bibitem{Mayr16}
C. Mayr et al., ``A biological-realtime neuromorphic system in 28 nm CMOS using low-leakage switched capacitor circuits,'' \textit{IEEE Transactions on Biomedical Circuits and Systems}, vol. 10, no. 1, pp. 243-254, 2016.%

\bibitem{Seo11}
J.-S. Seo et al., ``A 45nm CMOS neuromorphic chip with a scalable architecture for learning in networks of spiking neurons,'' \textit{Proc. of IEEE Custom Integrated Circuits Conference (CICC)}, 2011. 

\bibitem{Kim15}
J. K. Kim et al., ``A 640M pixel/s 3.65 mW sparse event-driven neuromorphic object recognition processor with on-chip learning,'' \textit{IEEE Symposium on VLSI Circuits (VLSI-C)}, pp.~C50-C51, 2015.

\bibitem{Akopyan15}
F. Akopyan et al., ``TrueNorth: Design and tool flow of a 65 mW 1 million neuron programmable neurosynaptic chip,'' \textit{IEEE Transactions on Computer-Aided Design of Integrated Circuits and Systems}, vol. 34, no. 10, pp. 1537-1557, 2015.%

\bibitem{Davies18}
M. Davies et al., ``Loihi: A neuromorphic manycore processor with on-chip learning,'' \textit{IEEE Micro}, vol. 38, no. 1, pp. 82-99, 2018.

\bibitem{Markram12}
H. Markram, W. Gerstner and P. J. Sj\"ostr\"om, ``Spike-timing-dependent plasticity: a comprehensive overview,'' \textit{Frontiers in Synaptic Neuroscience}, vol. 4, no. 2, 2012.

\bibitem{Brader07}
J. M. Brader, W. Senn and S. Fusi, ``Learning real-world stimuli in a neural network with spike-driven synaptic dynamics,'' \textit{Neural Computation}, vol. 19, no. 11, pp. 2881-2912, 2007.

\bibitem{Frenkel17a}
C. Frenkel et al., ``A fully-synthesized 20-gate digital spike-based synapse with embedded online learning,'' \textit{Proc. of IEEE International Symposium on Circuits and Systems (ISCAS)}, pp. 17-20, 2017.

\bibitem{Lin14}
P. Lin, S. Pi and Q. Xia, ``3D integration of planar crossbar memristive devices with CMOS substrate,'' \textit{Nanotechnology}, vol. 25, no. 40, ID~405202, 2014.

\bibitem{Rofeh15}
J. Rofeh et al., ``Vertical integration of memristors onto foundry CMOS dies using wafer-scale integration,'' \textit{Proc. of IEEE Electronic Components and Technology Conference (ECTC)}, pp. 957-962, 2015.

\bibitem{Indiveri10}
G. Indiveri, F. Stefanini and E. Chicca, ``Spike-based learning with a generalized integrate and fire silicon neuron,'' \textit{Proc. of IEEE International Symposium on Circuits and Systems (ISCAS)}, pp. 1951-1954, 2010.

\bibitem{Hodgkin52}
A. L. Hodgkin and A. F. Huxley, ``A quantitative description of membrane current and its application to conduction and excitation in nerve,'' \textit{Journal of Physiology}, vol. 117, no. 4, pp. 500-544, 1952.

\bibitem{Izhikevich03}
E. M. Izhikevich, ``Simple model of spiking neurons,'' \textit{IEEE Transactions on Neural Networks}, vol. 14, no. 6, pp. 1569-1572, 2003.

\bibitem{Brette05}
R. Brette and W. Gerstner, ``Adaptive exponential integrate-and-fire model as an effective description of neuronal activity,'' \textit{Journal of Neurophysiology}, vol. 94, no. 5, pp. 3637-3642, 2005.

\bibitem{LeCun98}
Y. LeCun and C. Cortes, ``The MNIST database of handwritten digits,'' 1998 [Online]. Available: \url{http://yann.lecun.com/exdb/mnist/}.

\bibitem{Mortara94}
A. Mortara and E. A. Vittoz, ``A communication architecture tailored for analog VLSI artificial neural networks: intrinsic performance and limitations,'' \textit{IEEE Transactions on Neural Networks}, vol. 5, no. 3, pp.~459-466, 1994.

\bibitem{Boahen00}
K. A. Boahen, ``Point-to-point connectivity between neuromorphic chips using address events,'' \textit{IEEE Transactions on Circuits and Systems II}, vol.~47, no. 5, pp. 416-434, 2000.

\bibitem{Cassidy11}
A. Cassidy, A. G. Andreou and J. Georgiou, ``A combinational digital logic approach to STDP,'' \textit{Proc. of IEEE International Symposium of Circuits and Systems (ISCAS)}, pp. 673-676, 2011.

\bibitem{Chen11}
G. Chen et al., ``A dense 45nm half-differential SRAM with lower minimum operating voltage,'' \textit{IEEE International Symposium on Circuits and Systems (ISCAS)}, pp. 57-60, 2011.

\bibitem{Arthur06}
J. V. Arthur and K. Boahen, ``Learning in silicon: Timing is everything,'' \textit{Proc. of Advances in Neural Information Processing Systems (NIPS)}, pp.~75-82, 2006.

\bibitem{Thorpe01}
S. Thorpe, A. Delorme and R. Van Rullen, ``Spike-based strategies for rapid processing,'' \textit{Neural Networks}, vol. 14, no. 6-7, pp. 715-725, 2001.

\bibitem{Yu13}
Q. Yu, et al., ``Rapid feedforward computation by temporal encoding and learning with spiking neurons,'' \textit{IEEE Transactions on Neural Networks and Learning Systems}, vol. 24, no. 10, pp. 1539-1552, 2013.

\bibitem{Qiao17}
N. Qiao, C. Bartolozzi and G. Indiveri, ``An ultralow leakage synaptic scaling homeostatic plasticity circuit with configurable time scales up to 100 ks,'' \textit{IEEE Transactions on Biomedical Circuits and Systems}, vol.~11, no. 6, pp. 1271-1277, 2017.

\bibitem{Sheik12}
S. Sheik, E. Chicca and G. Indiveri, ``Exploiting device mismatch in neuromorphic VLSI systems to implement axonal delays,'' \textit{Proc. of IEEE International Joint Conference on Neural Networks (IJCNN)}, 2012.

\bibitem{Frenkel17b}
C. Frenkel, J.-D. Legat and D. Bol, ``A compact phenomenological digital neuron implementing the 20 Izhikevich behaviors,'' \textit{Proc. of IEEE Biomedical Circuits and Systems Conference (BioCAS)}, pp. 677-680, 2017.

\bibitem{Soleimani12}
H. Soleimani, A. Ahmadi and M. Bavandpour, ``Biologically inspired spiking neurons: Piecewise linear models and digital implementation,'' \textit{IEEE Transactions on Circuits and Systems I: Regular Papers}, vol.~59, no.~12, pp.~2991-3004, 2012.

\bibitem{Yang15}
S. Yang et al., ``Cost-efficient FPGA implementation of basal ganglia and their Parkinsonian analysis,'' \textit{Neural Networks}, vol.~71, pp.~62-75, 2015.

\bibitem{Neftci13}
E. Neftci et al., ``Synthesizing cognition in neuromorphic electronic systems,'' \textit{Proceedings of the National Academy of Sciences}, vol. 110, no. 37, pp. E3468-E3476, 2013.

\bibitem{Koch99}
C. Koch, \textit{Biophysics of computation: information processing in single neurons}, Oxford university press, 1999.

\bibitem{Moon00}
S.-W. Moon, J. Rexford and K. G. Shin, ``Scalable hardware priority queue architectures for high-speed packet switches,'' \textit{IEEE Transactions on Computers}, vol. 49, no. 11, pp. 1215-1227, 2000.

\bibitem{Courbariaux16}
M. Courbariaux et al., ``Binarized neural networks: Training deep neural networks with weights and activations constrained to +1 or -1,'' \textit{arXiv preprint arXiv:1602.02830}, 2016.

\bibitem{Moons17}
B. Moons et al., ``Minimum energy quantized neural networks,'' \textit{arXiv preprint arXiv:1711.00215}, 2017.

\bibitem{Rueckauer17}
B. Rueckauer et al., ``Conversion of continuous-valued deep networks to efficient event-driven networks for image classification,'' \textit{Frontiers in Neuroscience}, vol.~11, p.~682, 2017.

\bibitem{Diehl15}
P. U. Diehl et al., ``Fast-classifying, high-accuracy spiking deep networks through weight and threshold balancing,'' \textit{Proc. of International Joint Conference on Neural Networks (IJCNN)}, 2015.

\bibitem{Whatmough17}
P. N. Whatmough et al., ``A 28nm SoC with a 1.2 GHz 568nJ/prediction sparse deep-neural-network engine with $>$0.1 timing error rate tolerance for IoT applications,'' \textit{Proc. of IEEE International Solid-State Circuits Conference (ISSCC)}, 2017.

\bibitem{Kreiser17}
R. Kreiser et al., ``On-chip unsupervised learning in winner-take-all networks of spiking neurons,'' \textit{Proc. of IEEE Biomedical Circuits and Systems Conference (BioCAS)}, pp. 424-427, 2017.

\bibitem{Benjamin14}
B. V. Benjamin et al., ``Neurogrid: A mixed-analog-digital multichip system for large-scale neural simulations,'' \textit{Proceedings of the IEEE}, vol.~102, no. 5, pp. 699-716, 2014.%

\bibitem{Park14}
J. Park et al., ``A 65k-neuron 73-Mevents/s 22-pJ/event asynchronous micro-pipelined integrate-and-fire array transceiver,'' \textit{Proc. of IEEE Biomedical Circuits and Systems Conference (BioCAS)}, 2014.%

\bibitem{Indiveri15b}
G. Indiveri, F. Corradi and N. Qiao, ``Neuromorphic architectures for spiking deep neural networks,'' \textit{Proc. of IEEE Electron Devices Meeting (IEDM)}, 2015.

\bibitem{Stromatias15}
E. Stromatias et al., ``Scalable energy-efficient, low-latency implementations of trained spiking deep belief networks on SpiNNaker,'' \textit{Proc. of IEEE International Joint Conference on Neural Networks (IJCNN)}, 2015.

\bibitem{Merolla14}
P. A. Merolla et al., ``A million spiking-neuron integrated circuit with a scalable communication network and interface,'' \textit{Science}, vol.~345, no.~6197, pp.~668-673, 2014.

\bibitem{Park17}
J. Park et al., ``Hierarchical address event routing for reconfigurable large-scale neuromorphic systems,'' \textit{IEEE Transactions on Neural Networks and Learning Systems}, vol. 28, no. 10, pp. 2408-2422, 2017.

\bibitem{Hoppner17}
S. H\"oppner et al., ``Dynamic voltage and frequency scaling for neuromorphic many-core systems,'' \textit{IEEE International Symposium on Circuits and Systems (ISCAS)}, 2017.

\bibitem{Partzsch17}
J. Partzsch et al., ``A fixed point exponential function accelerator for a neuromorphic many-core system,'' \textit{Proc. of IEEE International Symposium on Circuits and Systems (ISCAS)}, 2017.

\end{thebibliography}
\end{document}